\documentclass[sigconf]{acmart}

\usepackage{tikz}
\usepackage{filecontents}
\usepackage{amsmath,amsfonts}
\usepackage{graphicx}
\usepackage{textcomp}
\usepackage{xcolor}
\usepackage{algorithm}
\usepackage{algorithmic}
\usepackage{multirow}
\usepackage{booktabs}
\usepackage{adjustbox}
\usepackage{multirow}
\usepackage{epstopdf}
\usepackage{url}

\usepackage{mathrsfs}

\usepackage{bbm}
\usepackage{amsfonts}
\usepackage{array}
\usepackage{booktabs} % For formal tables
\usepackage{multirow}
\usepackage{colortbl}
\usepackage{color}
\usepackage{xspace}
\usepackage[flushleft]{threeparttable}
\usepackage{url}
\usepackage{array}
\usepackage{mdwmath}
\usepackage{mdwtab}
\usepackage{eqparbox}
\usepackage{subfigure,graphicx}
\usepackage{enumerate}
\usepackage{epstopdf}
\usepackage{epstopdf}
\usepackage[switch]{lineno}
\usepackage{diagbox}
\usepackage{wasysym}
\usepackage{pifont}
\usepackage{balance}
\usepackage{footnote}
\usepackage{graphicx}
\usepackage{xcolor}
\def\BibTeX{{\rm B\kern-.05em{\sc i\kern-.025em b}\kern-.08em
    T\kern-.1667em\lower.7ex\hbox{E}\kern-.125emX}}
\def\ie{\textit{i.e.}}
\def\etc{\textit{etc.}~}
\def\eg{\textit{e.g.}}

\newcommand{\tabincell}[2]{\begin{tabular}{@{}#1@{}}#2\end{tabular}}

\makeatletter
\def\hlinew#1{%
  \noalign{\ifnum0=`}\fi\hrule \@height #1 \futurelet
   \reserved@a\@xhline}
\makeatother

\def\bluea#1{{\color{black}#1}}%jpp blue
\def\reda#1{{\color{black}#1}}%zbl red
\def\graya#1{{\color{black}#1}}%need to change
\def\bluea#1{{\color{black}#1}}%jpp new 
\def\reda#1{{\color{black}#1}}%jpp new 
\def\bluej#1{{\color{black}#1}}%ccs major
\def\redj#1{{\color{black}#1}}%ccs major
\def\qi#1{{\color{black}#1}}%ccs major
\AtBeginDocument{%
  \providecommand\BibTeX{{%
    \normalfont B\kern-0.5em{\scshape i\kern-0.25em b}\kern-0.8em\TeX}}}

\copyrightyear{2021}
\acmYear{2021}
\setcopyright{none}
\acmConference[CCS '21]{Proceedings of the 2021 ACM SIGSAC Conference on Computer and Communications Security}{November 15--19, 2021}{Virtual Event, Republic of Korea}
\acmPrice{15.00}
\acmDOI{10.1145/3460120.3485383}
\acmISBN{978-1-4503-8454-4/21/11}

\begin{document}
\fancyhead{}

\title{Black-box Adversarial Attacks on Commercial Speech Platforms with Minimal Information}


\author{Baolin Zheng$^{1,*}$, \ Peipei Jiang$^{1,*}$, \ Qian Wang$^{1,\dagger}$, \ Qi Li$^{2}$, \ Chao Shen$^{3}$, \ Cong Wang$^{4}$, \ Yunjie Ge$^{1}$, \ Qingyang Teng$^{1}$, \ and Shenyi Zhang$^{1}$}
\affiliation{\institution{$^{1}$ School of Cyber Science and Engineering, Wuhan University} }
\affiliation{\institution{$^{2}$ Institute of Network Sciences and Cyberspace, Tsinghua University; BNRist}}
\affiliation{\institution{$^{3}$ School of Cyber Science and Engineering, Xi'an Jiaotong University}}
\affiliation{\institution{$^{4}$ Department of Computer Science, City University of Hong Kong}}
\affiliation{\institution{\{baolinzheng, ppjiang, qianwang\}@whu.edu.cn, qli01@tsinghua.edu.cn, chaoshen@xjtu.edu.cn, congwang@cityu.edu.hk, \{yunjiege, qingyangteng, shenyizhang\}@whu.edu.cn}}

\thanks{$^*$ The first two authors contributed equally to this work. \\ 
$^{\dagger}$ Qian Wang is the corresponding author.}

% \author{
%     Yifeng Zheng\city\thanks{The first two authors contributed equally to this work.}, Xingliang Yuan\city\footnotemark[\value{footnote}], Xinyu Wang\city, Jinghua Jiang\city\xian, Cong Wang\city, Xiaolin Gui\xian \\
% %The command \footnotemark[\value{footnote}] inserts a superscript corresponding to the current value of the counter footnote.
%     \affaddr{\city Department of Computer Science, City University of Hong Kong, Hong Kong}\\
%        \affaddr{\xian Department of Computer Science and Technology, Xi'an Jiaotong University, Xi'an, China}\\
%        \email{\{yifezheng2, xinywang, congwang\}@cityu.edu.hk,\{xinglyuan3-c, jjinghua2-c\}@my.cityu.edu.hk}\\
%        \email{xlgui@mail.xjtu.edu.cn}
% }

\begin{abstract}
Adversarial attacks against commercial black-box speech platforms, including cloud speech APIs and voice control devices, have received little attention until recent years. Constructing such attacks is difficult mainly due to the unique characteristics of time-domain speech signals and the much more complex architecture of acoustic systems. The current ``black-box'' attacks all heavily rely on the knowledge of prediction/confidence scores or other probability information to craft effective adversarial examples (AEs), which can be intuitively defended by service providers without returning these messages. In this paper, we take one more step forward and propose two novel adversarial attacks in more practical and rigorous scenarios. For commercial cloud speech APIs, we propose Occam, a decision-only black-box adversarial attack, where only final decisions are available to the adversary. In Occam, we formulate the decision-only AE generation as a discontinuous large-scale global optimization problem, and solve it by adaptively decomposing this complicated problem into a set of sub-problems and cooperatively optimizing each one. Our Occam is a one-size-fits-all approach, which achieves 100\% success rates of attacks (SRoA) with an average SNR of 14.23dB, on a wide range of popular speech and speaker recognition APIs, including Google, Alibaba, Microsoft, Tencent, iFlytek, and Jingdong, outperforming the state-of-the-art black-box attacks. For commercial voice control devices, we propose NI-Occam, the first non-interactive physical adversarial attack, where the adversary does not need to query the oracle and has no access to its internal information and training data. We, for the first time, combine adversarial attacks with model inversion attacks, and thus generate the physically-effective audio AEs with high transferability without any interaction with target devices. Our experimental results show that NI-Occam can successfully fool Apple Siri, Microsoft Cortana, Google Assistant, iFlytek and Amazon Echo with an average SRoA of 52\% and SNR of 9.65dB, shedding light on non-interactive physical attacks against voice control devices.
\end{abstract}



\maketitle

\section{Introduction}\label{sec:intro}
Nowadays, with the advance of speech and speaker recognition technologies, they are reshaping the way we interact with ubiquitous smart devices. More specifically, automatic speech recognition (ASR) technologies~\cite{graves2013speech} allow machines to understand human voices, while speaker recognition (SR) technologies~\cite{reynolds2002overview, DBLP:journals/prl/Furui97} enable machines to identify a person from the characteristics of his/her voices\footnote{To facilitate differentiation, in the following discussions we abbreviate automatic speech recognition and speaker recognition to ASR and SR, respectively.}. As a result, ASR and SR have become universal in our daily lives, ranging from personal voice assistants (PVAs)~\cite{Hoy2018Alexa,DBLP:journals/tochi/AmmariKTB19} to biometric authentication{~\cite{tao2010biometric,bhattacharyya2009biometric} }on various smart devices. The popularity of such speech services allows people to greatly enjoy the convenience of integrating speech as a new input for smart devices to perform daily and even complicated tasks. For example, Amazon has released Alexa~\cite{Alexa} and Auto SDK~\cite{Alexa_SDK} that allow users to make credit card payments and control vehicles with voice interaction, respectively. 

Despite their wide applications, the excessive use of voice commands in safety-critical systems, like autonomous driving{~\cite{geiger2012we,levinson2011towards}} and biometric identification, also poses potential safety hazards. A line of recent researches{~\cite{DBLP:conf/uss/CarliniMVZSSWZ16, DBLP:conf/ccs/ZhangYJZZX17,DBLP:conf/ndss/AbdullahGPTBW19} } have extensively demonstrated the vulnerability of acoustic systems to numerous types of abnormal audios, such as noises and inaudible ultrasounds. These attacks, however, can be easily detected and/or defended by differentiating and analyzing the nature (\textit{i.e.}, legitimate or malicious) of the received audio signals. Inspired by the resounding success of adversarial attacks against image recognition systems \cite{szegedy2013intriguing, goodfellow2014explaining, carlini2017towards, moosavi2016deepfool}, more recent researches have begun to investigate the feasibility of adversarial examples in the audio domain \cite{huAdversarial2019}, as shown in Table~\ref{tab:progresscomparison}.

The very first attempts made by Carlini {\em et al.}~\cite{CWaudio} and Yuan {\em et al.}~\cite{CommanderSong} have shown that ASR systems are also inherently vulnerable to audio AEs in the white-box scenarios, where the attackers can make full use of a prior knowledge of the structure and parameters inside the system. When it comes to black-box settings, however, the success of adversarial attacks in the image domain is hard to be ported to the audio domain, mainly owing to the multiple non-trivial challenges presented by the unique characteristics of time-domain speech signals and the much more complex architecture of acoustic systems.

{\color{blue}
\begin{table*}[!t]
%\footnotesize{}
\small
% \centering
%\tbl{An overview of recent progress of adversarial attacks in the audio domain.}
\caption{An overview of the state-of-the-art adversarial audio attacks against ASR and SR systems.}
\vspace{-2mm}
\label{tab:progresscomparison} %\vspace{-1mm}
    \begin{tabular}{c|c|c|c|c|c|c|c}
   % \toprule
    \hlinew{1.2pt}
    \textbf{Method} & \tabincell{c}{\textbf{Threat Scenario}} & \tabincell{c}{\textbf{Attack Type$^\ddag$}} & \textbf{Task}& \textbf{Knowledge}&\textbf{Commercial$^\S$}&\textbf{Queries}&\textbf{SRoA$^\dag$}\\
    %\midrule

    \hlinew{1.2pt}
    Carlini {\em et al.}~\cite{CWaudio}&White-box& Digital  & ASR &Gradient & \ding{55} &$\sim$1000&100\% \\
    \hline
    CommanderSong ~\cite{CommanderSong} &White-box& Digital  & ASR &Gradient& \ding{55}& $\sim$100 &100\% \\
    \hline
    Taori {\em et al.}~\cite{Taori}  & Black-box & Digital & ASR & Prediction score& \ding{55}&$\sim$300,000 & \textless 40\% \\
    \hline
    SGEA~\cite{TIFS-SGEA} & Black-box & Digital & ASR &  Prediction score& \ding{55}& $\sim$300,000 & 100\% \\
    \hline
    \multirow{2}{*}{Devil's Whisper~\cite{Whisper}}& \multirow{2}{*}{Black-box} &Digital & \multirow{2}{*}{ASR} & \multirow{2}{*}{Confidence score} & \multirow{2}{*}{\checkmark} & \multirow{2}{*}{$\sim$1500} & $\sim$50\%$^\flat$ \\
    \cline{3-3} \cline{8-8}
    & &Physical &  &  &  &  & \textless 100\% \\
    \hline
    \multirow{2}{*}{SirenAttack~\cite{du2019sirenattack}}  & \multirow{2}{*}{Black-box}   & \multirow{2}{*}{Digital}  & ASR & Gradient & \ding{55} & $\sim$1000 & 100\%\\
    \cline{4-8}
      &   &   & SR & Prediction score & \ding{55} & $\sim$7500 & 100\%\\
    \hline
    {FakeBob~\cite{DBLP:journals/corr/abs-1911-01840}} & {Black-box}   & {Digital}  & {SR} & Prediction score & \ding{55} & $\sim$5000 & 100\%\\
    \hline
    \multirow{3}{*}{Ours} & \multirow{3}{*}{Black-box}  & \multirow{2}{*}{Digital} & ASR & \multirow{2}{*}{Final decision}& \multirow{2}{*}{\checkmark}& $\sim$30,000 & \multirow{2}{*}{100\%}\\
    \cline{4-4} \cline{7-7}
     &  &  & SR& &  & $\sim$10,000 &\\
    \cline{3-8}
     & & Physical & ASR & None$^\sharp$ & \checkmark & 0 &  $\sim$50\% \\
    %\hlinew{1.2pt}
    %\bottomrule
    \hlinew{1.2pt}
    \end{tabular}

\begin{tablenotes}
\item\footnotesize{
Note that, (i) $\ddag$: ``Digital'' means that audio AEs are injected into target systems, while ``Physical'' means that audio AEs are played over-the-air.
%$\ddag$: ``Limited targeted'' means that the target phrases are limited to several selected phrases for a substitute model. 
(ii) $\S$: `` \checkmark'' means that the target model is the commercial platforms, otherwise `` \ding{55}''.
% $\sharp$: ``-'' means the information was not available in this work.
(iii) $\dag$: \textbf{SRoA} denotes the success rate of attack. It calculates the proportion of adversarial examples that can successfully attack target systems.
%while it was evaluated in Devil's Whisper using the success rate of commands, that is, the number of successful commands to the total number of the commands.
(iv) $\sharp$: Our physical attack focuses on the non-interactive setting where the adversary has no access to the oracle (the target model). 
(v) $\flat$: The SRoA of ``$\sim$50\%'' is calculated from our reproduced experiments on 5 digital speech APIs. Since the confidence scores are not available from Alibaba, iFlytek, and Tencent, we have omitted the score-related processing step in reproduced experiments of Devil's Whisper on these three speech APIs. Targeting Google TTS and Microsoft ASR (which return confidence scores), we did not omit the score-related steps from Devil's Whisper. That is, we followed the original version of Devil's Whisper when conducting the experiments on Google and Microsoft. More details about our reproduced experiment can be found at Table~\ref{tab:apisnr}.
}
\end{tablenotes}

\end{table*}%
}

As evidenced by~\cite{Taori}, Taori {\em et al.} combined genetic algorithms \cite{whitley1994genetic} and gradient estimation \cite{chen2017zoo}, which have been proved effective in the image domain \cite{alzantot2018genattack, bhagoji2018practical}, to carry out a black-box attack against open-source DeepSpeech{~\cite{deepspeech}} but with only a success rate of 35\%. Besides, considering that the attacker requires query access to the last layer (\textit{i.e.}, logits) of DNNs inside the DeepSpeech, such attacks become unrealistic when applied to the closed-source ASR systems. More recently, Chen {\em et al.}~\cite{Whisper} showed that when the confidence scores are publicly known, several commercial ASR systems are vulnerable to adversarial inputs, but with only a very limited number of target commands. Almost instantaneously, Du {\em et al.}~\cite{du2019sirenattack} and Chen {\em et al.}~\cite{DBLP:journals/corr/abs-1911-01840} presented the first black-box adversarial attacks against SR systems that both heavily relied on prediction scores. However, such kind of scores defined by the attackers or provided by speech service APIs may be useless to legitimate users, so service providers can easily hide these scores to defend against the above black-box attacks. Besides, there also exist a significant number of speech service APIs (\textit{e.g.}, Alibaba Short Speech Recognition API~\cite{Alibaba}, iFlytek Speech-to-Text API~\cite{iFlytek}) which do not return any intermediate information (\textit{e.g.}, confidence/prediction scores or other probabilities), except for the final decision results, \textit{e.g.}, final transcriptions in ASR systems and user-ids in SR systems.

\graya{From the practical perspective, the prior efforts are important but not satisfactory enough with respect to the minimum information as required by the adversary to launch a successful black-box attack. We may ask: ``\textit{is it possible to launch effective and practical audio adversarial attacks against commercial black-box speech platforms with the minimum information}?'' We are facing this opportunity already, more facing a big challenge mainly due to the extreme lack of information about the target model. 
%Fortunately, although they cannot be directly applied in the audio domain, several technologies have been explored very freshly in the image domain that shed light on the significance of decision-based attacks.
Specifically, the acoustic system, which involves non-linear feature extraction steps to cope with the intricate frequency feature changes in the time dimension, is much more complicated than the image processing system. Furthermore, a speech vector usually contains nearly one hundred thousand variables, far exceeding the hundreds or thousands of pixels in images, \textit{i.e.}, MNIST and CIFAR-10 are 28$\times$28 and 32$\times$32 respectively. As reported in~\cite{DBLP:conf/iclr/YangLCS19}, the explicit interdependencies among the massive number of variables significantly hinder the successful construction of audio AEs.}

{
\subsection{Our Works}
Generally speaking, there are mainly two types of black-box speech platforms. One is commercial Cloud Speech APIs that provide audio services to users, and the other is commercial Voice Control Devices, such as Apple Siri, which perform the speech-to-text task in the physical world. In this paper, we present two attack schemes, Occam\footnote{\bluej{Occam comes from the famous Occam's razor that plurality should not be posited without necessity, indicating that our attacks against speech platforms can be performed with minimal information.}}, a decision-only attack on cloud speech APIs, and NI-Occam, a non-interactive physical attack on voice control devices.

\noindent\textbf{Occam.} In our first design, we take one more step forward and focus on real-world threat scenarios where the adversary has access to an oracle (target model) which returns only its final decision. 
\reda{We propose Occam, a decision-based black-box adversarial attack against cloud speech APIs. We demonstrate that various commercial speech API services, such as Google Cloud Speech-to-Text, Alibaba Cloud Speech-to-Text, and Microsoft Azure Speech Service, are inherently vulnerable to audio AEs generated by our Occam, even if no internal information is exposed to the adversary. 
}

Our key idea of Occam is to formulate the decision-based black-box attack against smart acoustic systems as a discontinuous large-scale global optimization problem, on the basis of the final discrete decision (the attacker can only obtain) and a large number of optimization variables incurred by the speech. \bluej{Inspired by this observation}, we develop a novel technique called CC-CMA-ES, which applies a cooperative co-evolution (CC) framework to the powerful covariance matrix adaptation evolution strategy (CMA-ES), to solve the large and complex problem in the strictly black-box setting. More specifically, CC-CMA-ES first decomposes the complicated problem into a set of smaller and simpler sub-problems, and then uses CMA-ES to cooperatively optimize each one by modeling their local geometries. To improve the attack efficiency, we further propose an adaptive counterpart, which allows the sub-problem size and the decomposition strategy to self-adapt to the environmentally changeable evolution process.

We conduct extensive experiments to evaluate our attack capabilities on both speech and speaker recognition tasks, \bluej{and also compare it with five decision-based black-box methods to demonstrate the superiority of Occam}. We first craft audio adversarial examples against the local DeepSpeech model in the strictly black-box setting, achieving perfect success rates in both targeted and untargeted attacks. Then, we launch black-box adversarial attacks on a wide range of commercial speech-to-text API services, including Google, Microsoft, Alibaba, Tencent, and iFlytek, with success rates of 100\% \bluej{and an average SNR of 14.37dB}. Furthermore, we verify the attack effectiveness against commercial SR systems including Microsoft and Jingdong. It still achieves success rates of 100\% \bluej{and an average SNR of 14.07dB}. %\redj{ADD COMPARISON RESULT?}

\noindent\textbf{NI-Occam.} In our second design, we further probe the possibility of launching more rigorous and practical attacks on voice control devices, where the adversary still has no access to internal information and training data of the oracle, and does not even need to make queries to probe it. We, for the first time, propose a non-interactive physical attack, named NI-Occam, which successfully attacks many commercial voice control devices without any interaction. We show that NI-Occam works well in real-world attack scenarios, where audio AEs are played over-the-air. 

% Without sending a large number of queries that are expensive in time and money, 

Our key idea of NI-Occam is to combine adversarial attacks with model inversion attacks~\cite{DBLP:conf/ccs/FredriksonJR15, DBLP:journals/corr/abs-2010-04092, DBLP:conf/cvpr/ZhangJP0LS20}. More specifically, we make the attempt to recover the key parts of natural commands audio that are critical for speech recognition on the original example via the gradient information. Since these two audios are well blended together in model inversion process, it is difficult to be separated by human ears, which significantly hinders people from recognizing the malicious audio. Finally, our proposed NI-Occam can successfully fool Apple Siri, Microsoft Cortana, Google Assistant, iFlytek, and Amazon Echo with an average SRoA of 52\%. Human perception experiments further show that after being heard once, only 6.4\% auido AEs can be recognized as target commands by volunteers.

% Apple Siri, Microsoft Cortana, iFlytek and Google Assistant, with a 55\% success rate of attack and 10.04dB SNR.

We emphasize that our attacks have the following highlights: 1) \textit{Practicality}. They are able to attack commercial black-box platforms in the real-world scenarios without any prior knowledge; 2) \textit{Generality}. They are able to attack a wide range of commercial cloud speech APIs and voice control devices; 3) \textit{Effectiveness}. They are able to automatically and easily generate audio AEs with high success rates of attack.

\noindent\textbf{Contribution.} Our major contributions are summarized as follows.

\noindent $\bullet$ \textit{\bluea{Generic} black-box attacks with the minimum required information}. We present a novel decision-only audio adversarial attack, named Occam, under the strictly black-box scenario where the attackers can rely solely on the final decisions available in any application cases, and this is quite different from the state-of-the-art black-box adversarial attacks against commercial Cloud Speech APIs. In this sense, our attack strategy is the first one that can fool both commercial ASR and SR services, to our best knowledge.

\noindent $\bullet$ \textit{Effective attacks with a perfect success rate of attack}. We thoroughly evaluate our attack on a wide range of popular open-source and commercial (A)SR systems, including Google, Alibaba, Microsoft, Tencent, iFlytek, Jingdong, and DeepSpeech systems. %Experimental results extensively demonstrate that our attack is highly effective with a success rate of 100\% and can successfully generate stealthy audio AEs with a minimum SNR of 9.48 dB.
Extensive experiments demonstrate that our attack is highly effective with a success rate of 100\% and an average SNR of 14.23dB on commercial speech services,
\bluea{outperforming the state-of-the-art black-box attacks on commercial cloud speech APIs.}

\bluea{
\noindent $\bullet$ \textit{Practical physical attacks without any interaction}. We explore the possibility of generating audio AEs and playing them against the commercial voice control devices over-the-air. We thus for the first time propose a new non-interactive physical attack, named NI-Occam, which can successfully fool various voice control devices, including Apple Siri, Microsoft Cortana, Google Assistant, iFlytek and Amazon Echo, without any feedback information from the target devices. The experimental results show that our over-the-air attack can achieve an average success rate of 52\% with SNR of 9.65dB. This observation is shedding light on non-interactive physical attacks against voice control devices.
}
}

\section{Background}\label{sec:bac}
% In this section, we provide appropriate backgrounds for speech recognition, speaker recognition and adversarial examples, respectively.
\subsection{Speech Recognition}\label{sec:SR}
\begin{figure}[t!]
  \centering
  \includegraphics[width=0.85\linewidth]{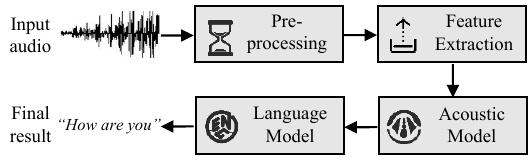} \\
  \caption{The architecture of a typical ASR system.}
  \vspace{-2mm}
  \label{fig:asr}
\end{figure}

Automatic speech recognition (ASR) systems allow machines to automatically convert speeches into texts, and have found tremendous applications. Typically, an ASR system consists of four main components: pre-processing, feature extraction, acoustic model, and language model, as in Figure~\ref{fig:asr}. Pre-processing plays an important role to filter out the frequencies beyond the range of human hearing and the segments below a specific energy threshold in the raw audio. Then, features are extracted via a feature extraction algorithm, such as Mel-frequency Cepstral Coefficients (MFCC)~\cite{DBLP:journals/corr/abs-1003-4083}, Linear Predictive Coefficient (LPC)~\cite{articlelpc}, Perceptual Linear Predictive (PLP)~\cite{articleplp},~\etc Different from image classification models \cite{Krizhevsky2012ImageNet} which take pixels as input, the acoustic model takes the extracted features as input, and outputs the probability of phonemes. In early ASR systems, Hidden Markov Model is one of the preferred techniques~\cite{DBLP:conf/naacl/AustinBCDKKMPRS89}. With the stupendous advance of deep learning, deep neural networks (DNNs)~\cite{DBLP:journals/spm/X12a} and especially recurrent neural networks (RNNs)~\cite{DBLP:conf/asru/RaoSP17, graves2013speech} have become the dominant choices today. An ASR system will finally produce the correct transcriptions conforming to grammatical rules via the language model.

\subsection{Speaker Recognition}
Due to the tremendous advance of DNN, speaker recognition (SR) systems are becoming increasingly popular in biometric authentication~\cite{DBLP:conf/icassp/LeiSFM14, DBLP:conf/interspeech/ZhangK17, DBLP:conf/icdcs/ChenRPWWWSM17}. The systems, which allow machines to correctly identify a person from his/her unique characteristics of voices, have various applications such as bank services and forensic tests.
%Similar to the fingerprint that provides an almost foolproof channel to identify the user owing to its  characteristics, voiceprint also has one-of-a-kind features to everyone.
The key step then is to extract users' voice features from a series of utterances which essentially consist of the underlying text information and the features of the speaker.

\begin{figure}[t!]
  \centering
  \includegraphics[width=0.9\linewidth]{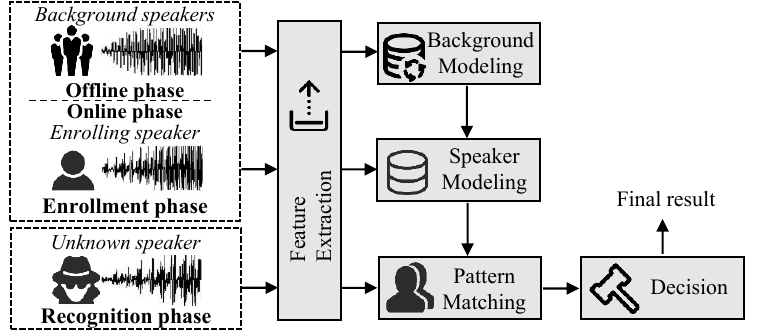} \\
  \caption{The architecture of a typical SR system.}
  \label{fig:srs}
  \vspace{-5mm}
\end{figure}

Figure \ref{fig:srs} provides the overview of a typical SR system \cite{DBLP:journals/speech/KinnunenL10}, which consists of two phases: enrollment and evaluation. Generally, in the enrollment phase, a set of background speakers are required to train the background model in the offline phase such that a speaker can provide a few utterances online to create a new specific speaker model. The technologies for generating such models can be generally summarized as three types: i-vector-PLDA \cite{DBLP:conf/interspeech/NandwanaFMCL19, DBLP:conf/icassp/NidadavoluIVD19}, GMM-UBM \cite{DBLP:journals/dsp/ReynoldsQD00, DBLP:journals/taslp/ReynoldsR95}, and DNN \cite{DBLP:conf/interspeech/ZhangK17, DBLP:conf/interspeech/SnyderGPK17}. During the evaluation phase, the unknown speaker's voice is taken as input and scored by the speaker models in the library. Based on the resulting scores, the decision model will generate the final recognition result.

There are two important sub-tasks in speaker recognition: speaker verification (SV)~\cite{DBLP:journals/dsp/ReynoldsQD00} and speaker identification (SI)~\cite{DBLP:journals/tc/Mohn71}. The former is to validate whether the current user is legitimate, \textit{i.e.}, output either $\textsf{accept}$ or $\textsf{reject}$. The latter aims to figure out the identity of the speaker among a set of enrolled ones. According to its text dependence, SR systems can also be divided into text-dependent and text-independent~\cite{reynolds2002overview}. The difference is that the text-dependent approach requires all speakers to utter pre-defined sentences. Despite higher accuracy, it is only used for the SV sub-task.

\subsection{Adversarial Examples}\label{sec:ae}
%With miraculous advances in deep learning, neural networks have been rapidly applied to a wide range of challenging areas, including image, speech and speaker recognition.
Despite their great success, the vulnerabilities of neural networks to adversarial examples have recently been extensively studied~\cite{szegedy2013intriguing, goodfellow2014explaining, carlini2017towards, moosavi2016deepfool}. An adversary can slightly revise the legitimate inputs to generate adversarial examples for fooling neural networks \cite{szegedy2013intriguing}. Generally speaking, there are two types of adversarial examples: untargeted and targeted ones, also known as dodging attacks and impersonation attacks, respectively.
Let $f(x) : x \in \mathcal{X} \rightarrow y \in \mathcal{Y}$ denote the recognition model that maps the input $x$ into the
corresponding output prediction $y$. Given the original input $x$ with the prediction $y$ such that $f(x) = y$, an untargeted adversarial example $x^*$ in the dodging attacks can be represented as:
\begin{equation}
    x^*, \quad s.t. \quad f(x^*) \neq y,~D(x, x^*) = \left\| x - x^{*}\right\|_p \leq \epsilon,
\end{equation}
where $D(x, x^*)$ is the distance between an original input $x$ and an untargeted adversarial example $x^*$, $\epsilon$ is a parameter used to limit this distance, and $p$ is typically 0, 2, or $\infty$. Similarly, in the impersonation attacks, for an original input $x$ and a specific $y^*$ such that $f(x) \neq y^*$, a targeted adversarial example $x^*$ can be represented as:
\begin{equation}
    x^*, \quad s.t.  \quad f(x^*) = y^*,~D(x, x^*) \leq \epsilon.
\end{equation}

\section{Occam: A Decision-only Digital Attack against Cloud Speech APIs}\label{sec:app}
\subsection{Threat Model}\label{sec:threatmodel}
\bluej{Nowadays, many commercial cloud speech platforms offer both ASR and SR API services,~\eg, Microsoft Azure Speech and Speaker Recognition Service API, and thus third-party developers can access commercial APIs if they have enrolled or paid for the services.  
The service providers, on the other hand, will not expose any parameters or the architecture of the target model because the internal information is commercially sensitive. Actually, a number of API services,~\eg, iFlytek, Alibaba, Tencent, and Jingdong, provide only the final decision results without exposing any other information. Therefore, it is important to explore a generic attack against both ASR and SR commercial APIs in this decision-based scenario. Note that although ASR tasks are different from SR tasks, we can treat them as the same problem in this design because the construction of audio AEs for ASR and SR APIs can be formulated as the same optimization problem. }

In this section, our target is commercial cloud speech services that open their APIs to the public, and we assume that the adversary intends to generate AEs against both ASR and SR services without any internal knowledge of the target model. More specifically, the adversary can only query the target model and obtain its final decision, which is a strict but more practical assumption in real-world applications. Considering the adversary's knowledge of the original audio, we make two different assumptions for the ASR and SR tasks respectively. For ASR systems, we assume that the adversary knows nothing about the dataset, and thus we utilize Text-to-Speech Service API to generate the audio of the target text. {For SR systems, the adversary only needs to collect the victim's one voice sample, which is readily available owing to the serious leakage of personal information,~\eg, public videos in social media.}

\subsection{Problem Formulation}
For an acoustic system, given a voice $x \in \mathbb{R}^{n}$ and a specific $y^*$ such that $f(x) \neq y^*$, a targeted AE $x^*$ can be described by
\begin{equation}\label{con:ori}
    x^*, \quad s.t.  \quad f(x^*) = y^*, D(x, x^*) \leq \epsilon.
\end{equation}
In the white-box setting, the problem can be formulated as
\begin{equation}\label{con:whitebox}
    \min \limits_{x^*} \mathcal{L}(x^*) = \mathcal{D}(x^*, x) + c \cdot \mathcal{J}(x^*, y^*),
\end{equation}
where $\mathcal{L(\cdot)}$ is the objective function, $\mathcal{J}(\cdot, \cdot)$ the loss function to check how well $x^*$ meets the adversarial requirement, and $c$ the adjustment parameter~\cite{CWaudio}. In the black-box setting without internal knowledge, we can reformulate it as an optimization problem as
\begin{equation}\label{con:targetattack}
    \min \limits_{x^*} \mathcal{L}(x^*) = \begin{cases}
        \mathcal{D}(x^*, x), & \text {if ${f(x^*) = y^*}$}, \\
        +\infty, & \text{otherwise.}
    \end{cases}
\end{equation}
Note that the $\mathcal{L}(x^*)$ is equal to $+\infty$ when $x^*$ is not adversarial. Thus, we try to find the adversarial region and minimize its distance from the original audio. Similarly, the construction of untargeted audio AEs in our decision-based black-box attack can be reformulated as
\begin{equation}\label{con:untargetattack}
    \min \limits_{x^*} \mathcal{L}(x^*) = \begin{cases}
        \mathcal{D}(x^*, x), & \text {if ${f(x^*) \neq f(x)}$}, \\
        +\infty, & \text{otherwise.}
    \end{cases}
\end{equation}

\begin{figure}[!t]
    \centering
    \subfigure[Continuous optimization]{
    \begin{minipage}[t]{0.48\linewidth}
        \centering
        \includegraphics[width=0.95\linewidth]{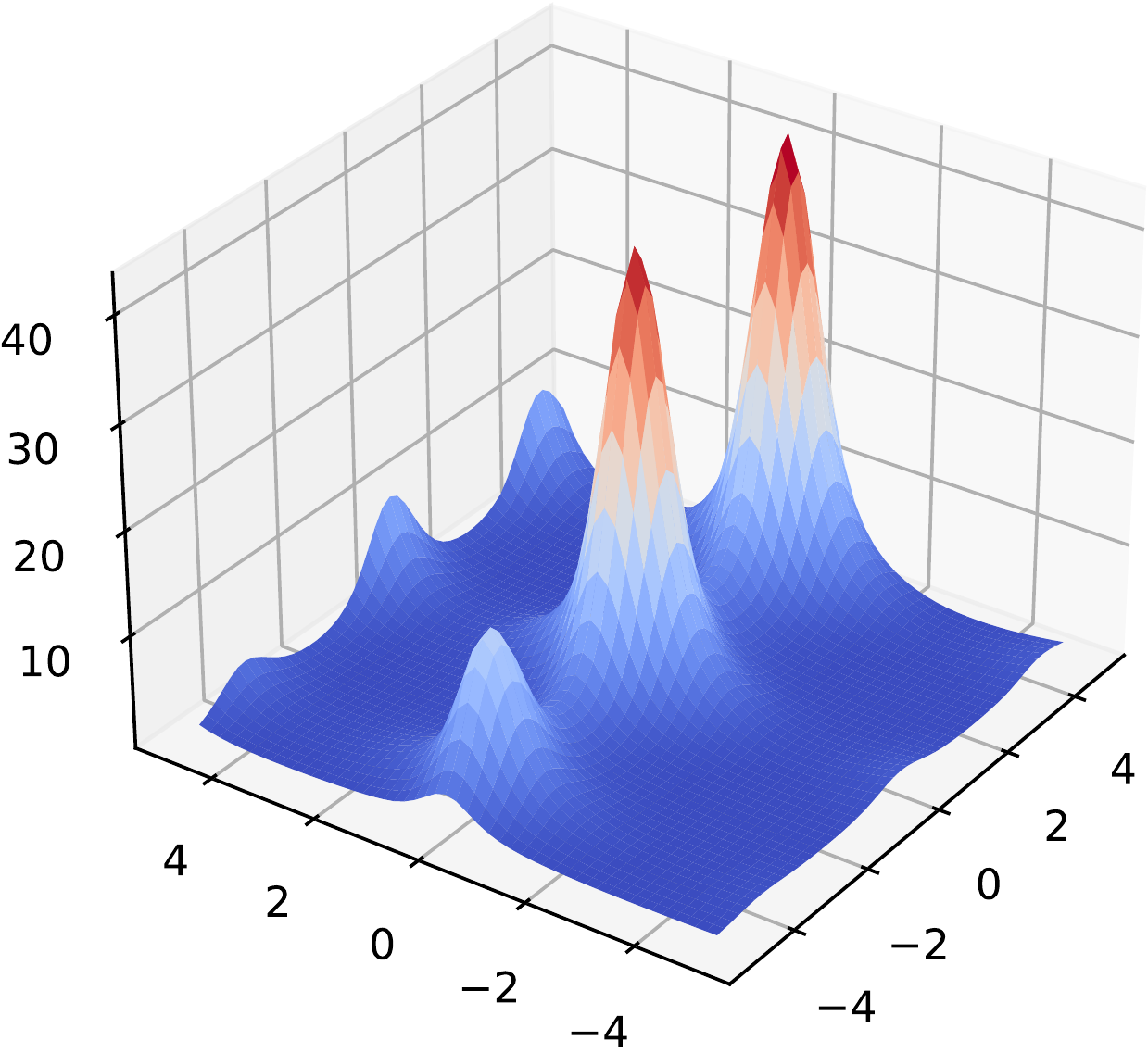}
        %\caption{Continuous optimization}
        \label{fig:successratebar}
    \end{minipage}}
    \subfigure[Discontinuous optimization]{
    \begin{minipage}[t]{0.48\linewidth}
        \centering
        \includegraphics[width=0.95\linewidth]{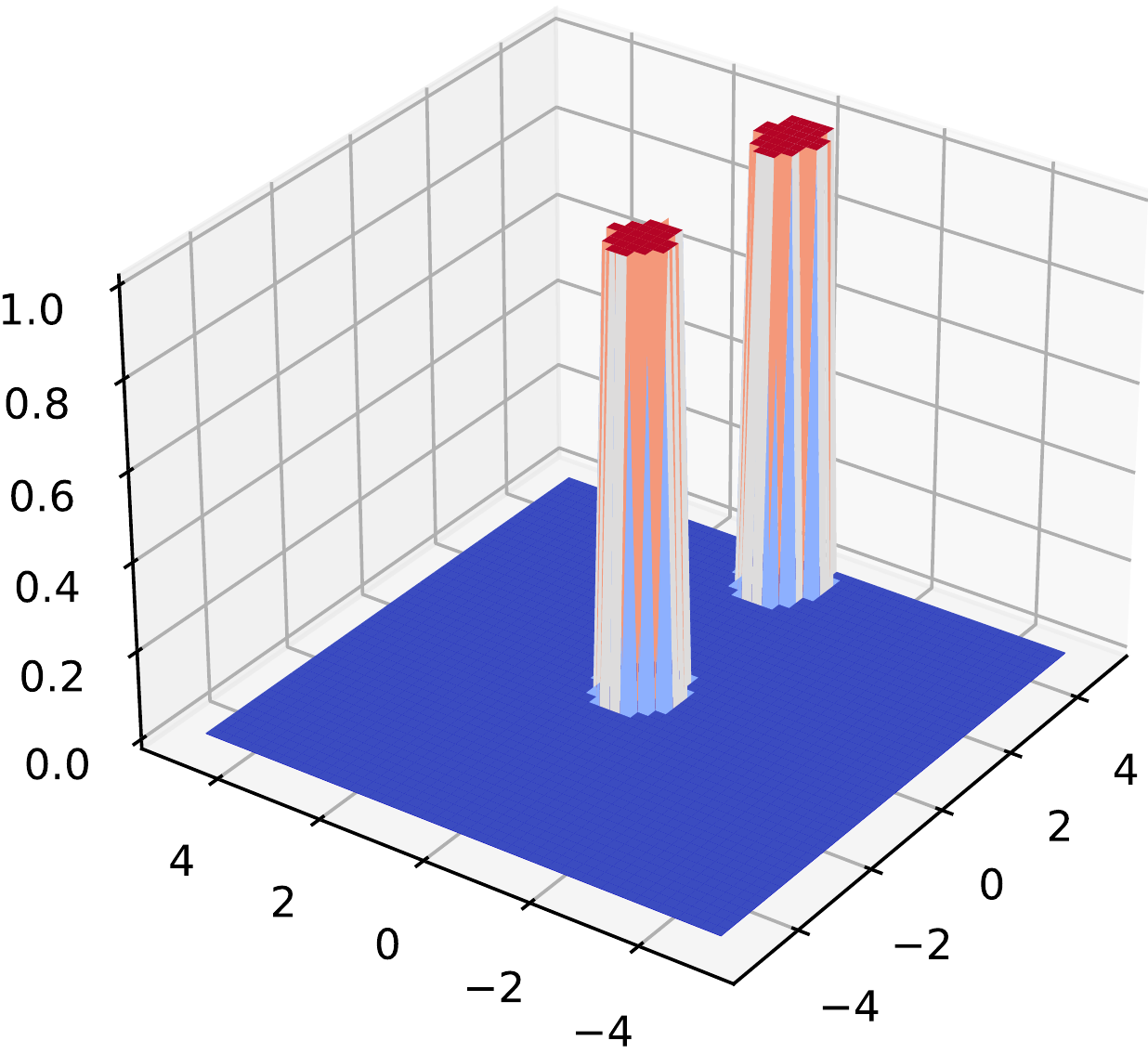}
        %\caption{Discontinuous optimization}
        \label{fig:querybar}
    \end{minipage}}
    \caption{An illustration of continuous (left) and discontinuous (right) optimization problem.}
    \label{fig:cpdp}
\end{figure}

\begin{figure}[t!]
  \centering
  \includegraphics[width=0.98\linewidth]{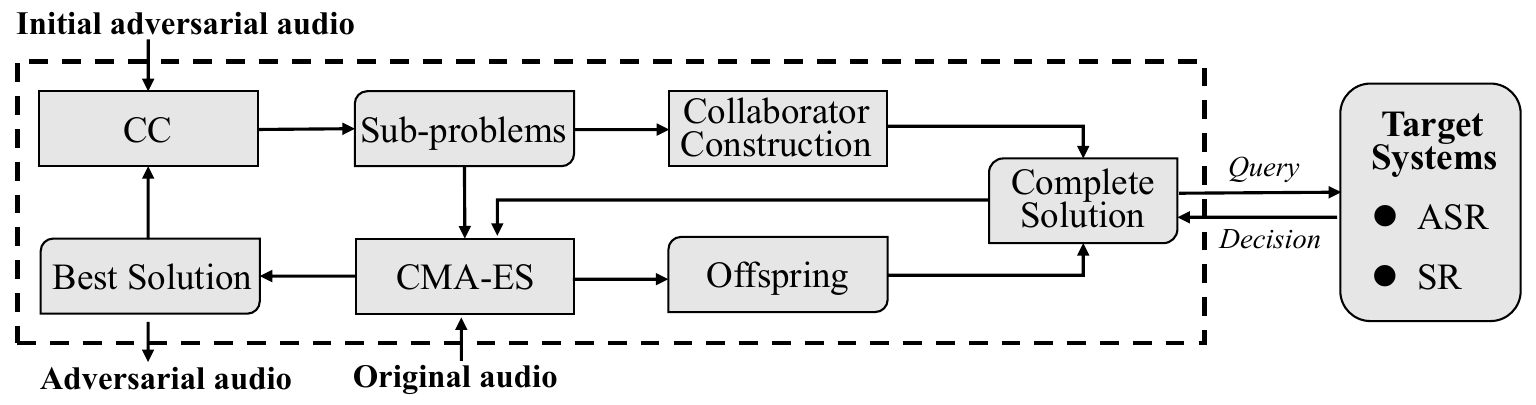} \\
  \caption{The architecture of our Occam.}
  \label{fig:attackarch}
  \vspace{-3mm}
\end{figure}

\subsection{Technical Challenges}
{As mentioned above, compared to the white-box and score-based settings, it is quite challenging to craft decision-based black-box adversarial examples against acoustic systems, since the attacker has no internal knowledge about the target model, except for the final decision corresponding to the query. Thus, we are facing several design challenges. First, depending on whether the output decision matches the adversarial target or not, the optimization space can be divided into a large non-adversarial region and a very small adversarial region. Hence, different from almost all previous adversarial voice attacks
%except for \cite{KAM2018}
which feature a continuous optimization problem, the construction of audio AEs in the strictly black-box scenario poses a difficult discontinuous optimization problem due to the extreme lack of information, as shown in Figure~\ref{fig:cpdp}. Second, the acoustic systems are extremely complicated, and we have to deal with intricate feature changes of the audios in the time dimension. Finally, because the audio sampling rate is very high (\textit{e.g.}, 16kHz), the large number of optimization variables in the speech vector presents another challenge, {\textit{i.e.}, the curse of dimensionality}, especially when there is a clear interdependence among the variables in audio AEs~\cite{DBLP:conf/iclr/YangLCS19}. The reason behind it is that as the number of optimization variables increases, the complexity of the problem grows exponentially, and the nature of the problem may also change. }

%(\textbf{what is the real challenge? to be more specific...})

\begin{algorithm}[!t]
    \caption{Occam}\label{alg:cc-cmaes}
    \begin{algorithmic}[1]
        \REQUIRE {The original audio $x$, the initial adversarial audio $x'$, the input space dimension $n$, the attack objective function $\mathcal{L(\cdot)}$, the binary search time $b$ and the total number of queries $T$.}
        \ENSURE The adversarial audio sample $x^*$.
        \STATE Initialize the following parameters:
        \begin{itemize}
        \item Covariance matrix $C = {I}_n$, $x^{*} = x'$, $t = 0$;
        \item $\delta, \mu, c_c, c_{cov} \in \mathbb{R}_+, m, \lambda \in \mathbb{Z}_+$ {(See Section~\ref{sec:parameter})}.
        \end{itemize}
        \STATE \reda{Query $b$ times to find the new adversarial point $x^*$ close to the decision boundary via binary search algorithm;}
        \STATE $t = t + b$;
        \STATE Choose a grouping strategy and a group number $m$ according to the adaptive scheme {(See Section~\ref{sec:adaptive})};
        \STATE Decompose the audio vector into $m$ disjoint parts with each part having $s = n/m$ dimensions; 
        \STATE Set $sub = 1$ to start an optimization cycle;
        \STATE {Extract two $s \times 1$ vectors $x^*_{sub}$ and $x_{sub}$ and an $s \times s$ covariance matrix ${C}_{sub}$ from $x^*$, $x$ and $C$, respectively;}
        \FOR {$i$ = $0$ to $\lambda$}
            \STATE Sample $z$ $\sim$ $\mathcal{N}(0, \sigma^{2} \cdot C_{sub})$;
            \STATE Generate one complete solution $\emph{solu}$ using $x^*_{sub}+\mu(x_{sub}-x^*_{sub})+z$ and collaborative information from other subspaces;
            \IF {$\mathcal{L}(\emph{solu})$ $\textless$ $\mathcal{L}(x^*)$}
                \STATE $x^*_{sub} =  x^*_{sub}+\mu(x_{sub}-x^*_{sub})+z$;                \STATE {Update $C_{sub}$ according to Eqs.~(\ref{con:formula2}) and (\ref{con:formula3});}
            \ENDIF
        \ENDFOR
        \STATE $t = t + \lambda$;
        \STATE {Use $x^*_{sub}$ and $\ C_{sub}$ to update $x^*$ and $C$, respectively;}
        \IF {$t \geq T$}
            {\RETURN {$x^*$.}}
        \ELSIF {$sub \textless m$}
            {\STATE $sub++$ and go to step 5;}
        \ELSE
            {\STATE Go to step 2;}
        \ENDIF
    \end{algorithmic}
\end{algorithm}

\subsection{Our Method}
As noted in our threat model, the lack of internal knowledge (\textit{e.g.}, structures, parameters, gradients, and scores) about the target model further exacerbates the difficulty of crafting AEs. All the adversary can do is to send a limited number of queries to probe the system as far as possible, and obtain the corresponding final decisions. More specifically, the initiation of our decision-based black-box adversarial attack only needs the final decision,~\eg, final transcription, and this kind of audio AE generation can be formulated as a discontinuous and large-scale global optimization problem. To solve it, we resort to the large-scale black-box optimization approach.

\noindent\textbf{Design Overview.} {Note that we are going to fool both commercial ASR and SR services with the minimum required information from the target model. 
To address the challenges, we propose a new class of cooperative co-evolution methods to generate effective audio AEs. \reda{Our method is mainly developed from CC-CMA-ES~\cite{DBLP:conf/ideal/LiuT13}. However, we cannot directly apply CC-CMA-ES to constructing audio AEs. 
In the audio domain, the correlations between variables will change
in the dynamic evolution process, which is not considered in~\cite{DBLP:conf/ideal/LiuT13}. To solve this problem, we devise an adaptive scheme to make our strategy self-adapt to the environmentally
changeable evolution process, the core of our cooperative
co-evolution framework.} 

In the literature, the CMA-ES~\cite{DBLP:journals/corr/Hansen16a}, known as an efficient evolutionary algorithm, has already demonstrated its good performance on many problems. But it will lose its effectiveness when applied to large-scale global optimization problems due to ``the curse of dimensionality''. A dimensionality reduction strategy, which uses the bilinear interpolation method to project the original space ($112\times112\times3$) to a lower-dimensional search space (\textit{e.g.,} $45\times45\times3$), has been carefully devised to effectively create adversarial images against face recognition models~\cite{CVPR}. However, this method is also not applicable to the audio domain. This is because bilinear interpolation works in two directions on images, while the inputs to the commercial Cloud Speech APIs are one-dimensional vectors from speeches. Thus, we for the first time introduce the general cooperative co-evolution (CC) framework to construct audio AEs in the strictly black-box setting as shown in Figure \ref{fig:attackarch}. Our CC framework can scale up CMA-ES to deal with large-scale optimization problems we are facing, by decomposing these challenging problems into a set of simpler and smaller sub-problems and cooperatively optimizing each of them. Considering that the group size and the decomposition strategy play a crucial role in CC, we further propose an adaptive scheme to improve the attack efficiency by letting the size of sub-problems and the decomposition strategy self-adapt to the environmentally changeable evolution process.}

Our Occam is presented in Alg.~\ref{alg:cc-cmaes}. {In each optimization cycle, the original problem is decomposed into a set of smaller and simpler subproblems according to the selected grouping strategy. Then, a new offspring is generated from the current solution in each subproblem by a subspace CMA-ES whose parameters are extracted from a global CMA-ES, and the complete candidate solution can be further obtained by using the collaborative information from other subproblems and the generated offspring.} The objective function is further used to evaluate the two solutions and choose the better one, based on which we update the covariance matrix accordingly. Next, we will describe each step of the algorithm in detail.

% \vspace{-10pt}

\subsubsection{Initialization}
As shown in Eq.~(\ref{con:targetattack}), the optimization routine should start from an adversarial point, because the value of the objective function is equal to $+\infty$ when the input is not adversarial. We first initialize $x^*$ with a natural adversarial sample distant from the original audio. More specifically, we utilize the text-to-speech API service to synthesize the desired speech as an initial adversarial audio sample against ASR systems. For SR systems, we initialize $x^*$ using an audio of the target speaker, which can be obtained from the speaker's self-recorded songs and videos posted on public social networks. Since the initial input audio is adversarial and the original one is not adversarial, we can first utilize the binary search algorithm to effectively approach the decision boundary.

\begin{figure}[t!]
  \centering
  \includegraphics[width=0.98\linewidth]{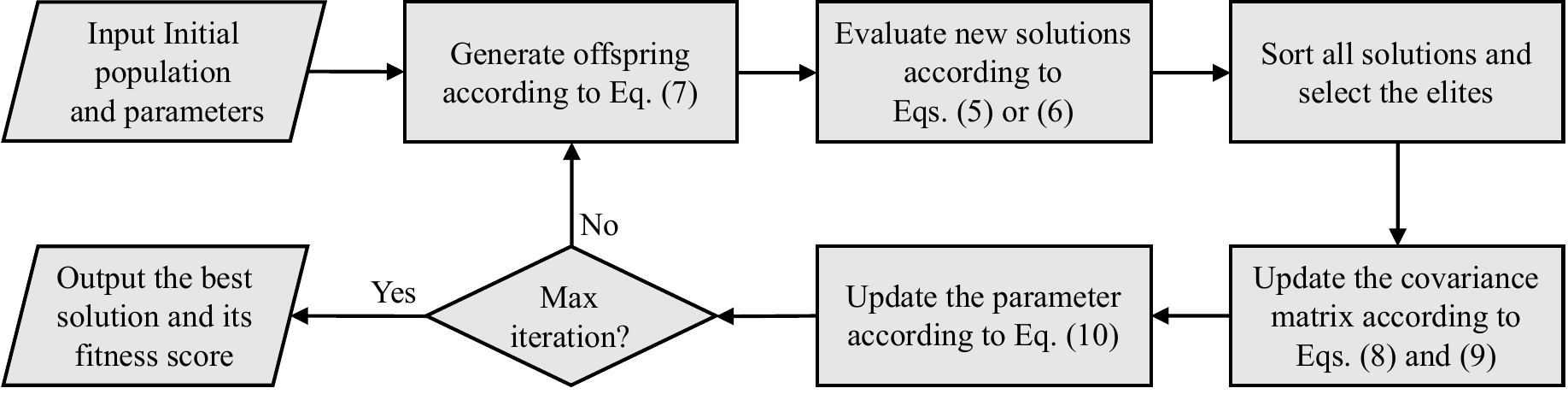} \\
  \caption{The diagram of CMA-ES.}
  \label{fig:cma-es-process}
\end{figure}

\subsubsection{Covariance Matrix Adaptation Evolution Strategy}\label{sec:cmaes}
\reda{CMA-ES, known as an efficient derivative-free method, generates offsprings by sampling a multivariate normal distribution with covariance $C$,~\ie, $\mathcal{N}(0, C)$. \bluea{To facilitate understanding, we provide a brief summary of CMA-ES, as shown in~Figure~\ref{fig:cma-es-process}.} Since the covariance is the measure of the relationship between two random variables, it can use the selected sample distribution to estimate the covariance for learning dependencies between variables~\cite{DBLP:journals/corr/Hansen16a}. Due to the unreliability of this estimation for small samples, the covariance matrix adaptation using the history information and the evolution path $P$, a sequence of successive and normalized steps, thus has been introduced. By adaptively updating the estimated covariance, CMA-ES is able to find better search directions, and achieve a powerful local search by modeling local geometries. 
}

\reda{Our framework uses a simple yet effective variant of CMA-ES,~\ie, (1+1)-CMA-ES~\cite{DBLP:conf/gecco/IgelSH06}. This version generates one candidate solution from current solution by sampling a random noise, and selects a better solution according to its objective function. Since the direction of optimization in our attack is to find a new adversarial audio closer to the original audio according to the objective function $\mathcal{L}(\cdot)$, we adopt a modified (1+1)-CMA-ES~\cite{CVPR} to improve its efficiency by adding a bias term $\mu(x - x^*)$ to current solution $x^*$ as
\begin{equation}\label{formula1}
x_{i+1}^{*}\sim \mathcal{N}(x_{i}^*+\mu(x - x_{i}^*), {\sigma}^2 \cdot C),
\end{equation}
where $\sigma$ is the global step size, $C$ is the covariance matrix that determines the shape of the distribution, and $\mu$ is a parameter that controls the degree of proximity towards the original audio. Furthermore, we can directly remove candidate solutions that are farther from the original audio, regardless of whether it is adversarial. 
}

Note that the covariance matrix $C$ plays a very important role in CMA-ES since it models the local geometries to improve the local search efficiency. However, the adaptation of covariance matrix with the complexity of $\mathcal{O}(n^3)$ may be infeasible when the input dimension $n$ is huge. To speed up the computation, the covariance matrix $C$ can be \reda{simplified} as a diagonal matrix \cite{CVPR} and updated adaptively by the evolution path $P$ as
\begin{equation}\label{con:formula2}
P=(1-c_c)P+\sqrt{c_c(2-c_c)}\frac{z}{\sigma},
\end{equation}
\begin{equation}\label{con:formula3}
C=(1-c_{cov})C+c_{cov}P(P)^T,
\end{equation}
where $c_c$ and $c_{cov}$ are the parameters controlling the adaptation of $P$ and $C$, respectively. The update enlarges the variance along the past successful directions for future search.

\subsubsection{Cooperative Co-evolution}\label{sec:CC}
It has been proven that the performance of evolution algorithms may drop significantly~\cite{DBLP:conf/ideal/LiuT13, DBLP:conf/cec/YangTY08a}, as the dimensionality of the problem increases, because the complexity of the problem grows exponentially and the property of the problem may also change. To scale up CMA-ES to the high-dimensional optimization problem, we use cooperative co-evolution (CC) to conduct the large-scale black-box optimization in a divide-and-conquer manner, by decomposing the large-scale problem into several smaller sub-problems and optimizing each sub-problem alternately and iteratively. Considering that each subproblem is only a part of the original problem, the collaborative information from other sub-problems is required to evaluate individuals in the current sub-problem. Generally speaking, the best solutions of each sub-problem in the current cycle are used as the collaborative information. However, considering the query limitation in our attack, we propose a greedy strategy to produce and update the collaborative information in our attack design. More specifically, we do not optimize these sub-problems concurrently. Instead, we optimize them alternatively and iteratively. Therefore, when optimizing a subproblem, we can evolve its values of the variables, which are used to replace those related to the current subproblem in the best solution, and generate a complete solution. The solution is further evaluated by the objective function $\mathcal{L}(\cdot)$ to locate a better solution. After the optimization of each subspace, the collaborative information will be updated correspondingly. \bluej{Since CC is a general framework based on the divide-and-conquer strategy for solving large-scale black-box optimization problems, it is generalizable to other black-box methods that are also trapped in these problems, and CC can also improve their effectiveness.}

\subsubsection{Adaptive Decomposition}\label{sec:adaptive}
%Since the group strategy that determines how to assign variables to different groups plays a crucial role in CC, various group strategies are presented.
The grouping strategy, which determines how to assign variables to different groups, plays a crucial role in CC.
However, there is insufficient knowledge about the correlations between variables, making manually devising or choosing the most suitable grouping strategy extremely hard when applying CC. Therefore, we propose an adaptive approach, which puts several popular grouping strategies into a candidate pool, and adaptively selects a proper decomposition strategy from it. The candidate pool includes four grouping strategies: \textit{Static grouping} (SG), \textit{Random grouping} (RG), \textit{Min-variance grouping} (MiVG) and \textit{Max-variance grouping} (MaVG). We adopt SG to preserve the information in the time domain, and RG~\cite{DBLP:conf/cec/YangTY08a} can help randomly allocate variables to subspaces for improving the probability of placing two interacting variables in the same subspace.
Considering that the covariance matrix is used to model the local geometries of the search directions, thus we adopt MiVG and MaVG, which are devised for CC-CMA-ES. Actually, they can minimize or maximize the diversity of the diagonal values of the variables in the same subspace.
%Considering that the covariance matrix is used to model the local geometries of search directions, MiVG and MaVG that are devised for CC-CMA-ES are also added to the candidate pool, which minimizes/maximizes the diversity of the diagonal values of the variables in the same subspace.
At the beginning of each optimization cycle, we randomly select a decomposition strategy. Based on its performance, we calculate the selection probabilities of each grouping strategy for the next cycle.
%In the next cycle, we will select a certain grouping strategy according to the selection probabilities.

We observe that the levels of interdependency among variables in the audio vector will change significantly from the natural audio to the audio AE during the optimization process~\cite{DBLP:conf/iclr/YangLCS19}. Therefore, we propose to adaptively adjust the group size to capture different interdependency levels in the dynamic evolution process.
Then, we can make a good trade-off between the effectiveness and the efficiency of the optimization. The details of the adaptive decomposition algorithm can be seen in Appendix \ref{sec:AdaptiveDecomposition}.

\subsubsection{Parameter Adjustment}\label{sec:parameter}
Our algorithm contains many hyper-parameters, such as $ \delta$, $\mu$, $\lambda$, $c_c$, $c_{cov}$ and $b$. Following~\cite{CVPR}, we set $\delta=0.001\cdot D({x}^{*},x)$, $c_c=0.01$, $c_{cov}=0.001$ and $b = 15$. We further set $\lambda = 30$ and $\mu = 0.08$, because $\mu$ has an important impact on the search process, we need to carefully tune $\mu$. Finally, we adopt the $1/5$th success rule \cite{DBLP:conf/gecco/Auger09} to update $\mu$ as
\begin{equation}\label{con:rule}
     \mu = \begin{cases}
        1.5\mu, & \text {if a better solution is obtained,} \\
        1.5^{-1/4}\mu, & \text{otherwise.}
    \end{cases}
\end{equation}

\section{NI-Occam: A Non-interactive Physical Attack against Voice Control Devices}
\subsection{Threat Model}
\bluea{In this section, our target is commercial voice control devices. We consider the most rigorous and practical assumption, called non-interactive physical setting, where the adversary makes no query to the oracle. Compared to prior physical attacks, the key advantage of non-interactive physical attacks is that we do not need to query the target devices for effective audio AE generation, thus saving the potentially large query cost. In this sense, this attack is the most practical one in the real world.}

\subsection{Technical Challenges}\label{sec:nichallenges}
\bluea{Compared to our decision-only adversarial attacks, non-interactive black-box setting is much more challenging since it further breaks the dependence on the final decision in the decision-only black-box scenario. That is to say, the target model is completely unknown to the adversary. Moreover, voice control devices also present additional challenges that the constructed audio AEs should remain robust even if they are played in the physical world. By physical attacks against voice control devices, we mean that audio AEs are played by a speaker and recorded by the device. \bluej{Since the effectiveness of audio AEs is greatly affected by the reverberation of the environment, and perturbations from the speaker and the microphone~\cite{DBLP:conf/ijcai/YakuraS19, schonherr2019robust}, it is really difficult to launch physical attacks.} These two obstacles pose severe challenges to craft effective audio AEs.}

% by perturbations from the transmission channel through air, the speaker, and the microphone
\subsection{Our Method}\label{sec:physical}

\bluea{As described in our threat model, the adversary will not issue any queries to probe the target model and obtain no feedback in the non-interactive black-box setting. Thus, our Occam cannot be applied in this case. Intuitively, it is almost impossible to directly construct an audio adversarial example against the target model by solving the optimization problem without any interaction. In fact, we are facing the problem of generating AEs with no information during the whole attacking process. In the image domain, there have been works that demonstrated AEs crafted for the target model are able to attack other unknown models, which is called the transferability of AEs. However, in the audio domain, the poor transferability of audio AEs~\cite{CommanderSong, du2019sirenattack} among different ASR systems indicates that we cannot directly leverage the transferability property especially when there are no interactions with the target model.}

\bluea{Observing that the ultimate goal of speech recognition systems is to perform the task of converting natural speech into text, we believe that the inclusion of the characteristics of natural command audios in the constructed audio AEs may help improve their transferability. Inspired by model inversion attacks~\cite{DBLP:conf/ccs/FredriksonJR15, DBLP:journals/corr/abs-2010-04092, DBLP:conf/cvpr/ZhangJP0LS20} that aim to recover input data or its sensitive attributes via the model output, we propose NI-Occam to craft audio AEs, where the command voice is recreated and implicitly embedded in the original music via the gradient update. The main reason behind is that it is hard for people to perform speech separation on our constructed audio AEs and further recognize the malicious speech commands. }

\bluea{Finally, audio AEs we constructed, just like natural command audios, will remain robust in the over-the-air attack and naturally effective in the physical world. Besides, compared to cloud speech APIs, voice control devices are more vulnerable to audio AEs containing command audios since they are more sensitive to voice commands than speech APIs, which has been demonstrated in Devil's Whisper~\cite{Whisper}. Thus, we propose NI-Occam, which realizes non-interactive physical attacks against voice control devices. Notably, this is the first physical attack that can effectively create audio AEs without any feedback information.
Previous physical attacks,~\eg, Devil's Whisper and FakeBob, rely much on returned scores to generate effective audio AEs, while our attack is a non-interactive one,~\ie, requiring no access to the target devices.}

\begin{figure}[t!]
  \centering
  \includegraphics[width=0.98\linewidth]{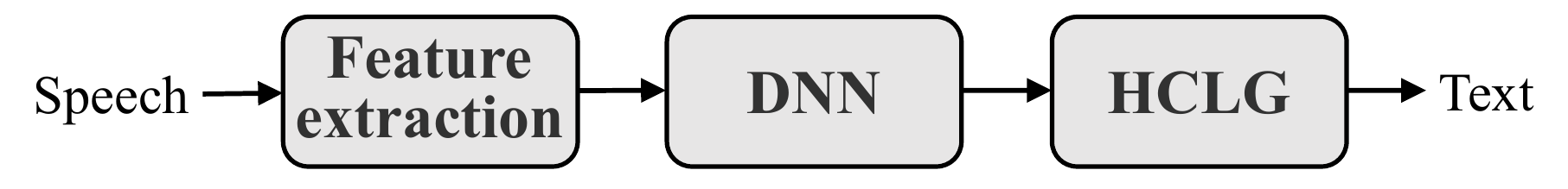} \\
  \caption{The main architecture of the Kaldi model.}
   \vspace{-3mm}
  \label{fig:Kaldi model architecture}
\end{figure}

\bluea{Our proposed NI-Occam is presented in Alg.~\ref{alg:gradientinversion}. To be specific, we choose the open-source Kaldi model (ASpIRE Chain Model) as the substitute model and the inversion model due to its simple structure of the neural network and the excellent performance. In Figure \ref{fig:Kaldi model architecture}, it can be seen that the Kaldi model obtains MFCC feature through feature extraction and takes these features as the input of DNN, while the output of DNN is the probability density function (pdf). According to the idea of ``pdf-id sequence matching'' proposed in CommanderSong~\cite{CommanderSong}, we can recover the command audios from their pdf-id sequences via the gradient inversion. Different from model inversion attacks that start from the Gaussian noise $z\sim\mathcal{N}(0, \sigma^2)$, adversarial attacks start from the original example $x$. Therefore, we add Gaussian noise $z$ into the original example $x$ to address this problem. The audio AEs can be described as
\begin{equation}\label{eq:gn}
x^*= \arg\min_{x^*} \mathcal{J}(z+x^*, y) \quad s.t. \left\|x-x^*\right\|_{\infty} < \epsilon,
\end{equation}
where $y$ is the probability value of the target pdf-id sequence, $\mathcal{J}(\cdot, \cdot)$ is the loss function~\cite{CommanderSong}. Note that, the Gaussian noise $z$ we added to the original example $x$ is very large. This is to alleviate the impact of the original example on the model inversion process. To facilitate convergence, we gradually attenuate the size of Gaussian noise $z$ in the iterative process. Besides, very recent works~\cite{DBLP:journals/corr/abs-2102-03726, DBLP:journals/corr/abs-2101-09930} have shown that the AdaBelief method  \cite{DBLP:conf/nips/ZhuangTDTDPD20} is beneficial to improve the transferability of adversarial examples, so we use the AdaBelief optimizer to solve Eq. (\ref{eq:gn}). The details of implementation are: the learning rate $\alpha$ is set to 0.003, the standard deviation $\sigma$ is set to 0.25, and the size of perturbation $\epsilon$
is set to 0.3.}

\begin{algorithm}[!t]
    \caption{NI-Occam}\label{alg:gradientinversion}
    \begin{algorithmic}[1]
        \REQUIRE {The original example $x$, the command audio $x'$, the loss function $\mathcal{J(\cdot, \cdot)}$, the Kaldi model $f$, the learning rate $\alpha$, the standard deviation $\sigma$ and the size of perturbation $\epsilon$.}
        \ENSURE The produced audio adversarial examples.
        \STATE Obtain the target pdf-id sequence $y$ through the Kaldi model $f$ and the command audio $x'$;
        \STATE Initialize $x^*$ with $x$, $X^*$ with $\emptyset$, and the learning rate of the AdaBelief Optimizer with $\alpha$;
        \WHILE {not converged yet}
            \STATE Sample $z$ $\sim$ $\mathcal{N}(0, \sigma^{2})$;
            \STATE Use the AdaBelief Optimizer to minimize $\mathcal{J}(z+x^*, y)$ and update $x^*$;
            \STATE Clip $x^*$ into the $\epsilon$ vicinity of $x$;
            \STATE $\sigma = 0.998 \times \sigma$;
            \STATE $X^*$ = $X^* \cup x^*$;
        \ENDWHILE
        \RETURN $X^*$.
    \end{algorithmic}
    
\end{algorithm}

\section{Implementation and Evaluations}
%We implement our attacks in Python, and evaluate the performance of the attacks against open-source ASR systems, speech-to-text API services, and speaker systems. We compare our attacks with the state-of-the-art gray-box and black-box attacks from the aspects of effectiveness and imperceptibility. %To further assess the generality of our attack, we also conduct experiments on images.

\subsection{Experiment Settings}\label{sec:settings}

\noindent\textbf{Experiment Design.}
\bluea{
To evaluate the performance of Occam,~\ie, our decision-only digital attacks on cloud speech APIs, 
we design four sets of experiments: attacks on open-source ASR systems, attacks on ASR services, attacks on SR services, and human perception of the audio AEs\footnote{The evaluations of attacks on open-source ASR systems and human perception are presented in Appendixes~\ref{app:deepspeech} and~\ref{app:human}, respectively}.}
For the attacks on ASR services, we choose ten frequently-used voice commands\footnote{
The target transcripts include \textit{call my wife, make it warmer, navigate to my home, open the door, open the website, play music, send a text, take a picture, turn off the light, and turn on the airplane mode}.
} 
that are expected to be recognized by the target systems. The original audios are selected from three datasets including Common Voice~\cite{commonvoice}, Song~\cite{CommanderSong}, and LibriSpeech~\cite{panayotov2015librispeech}. 
We select ten test samples in each experiment to evaluate the performance. To generate audio AEs, we need to query the target commercial speech-to-text cloud services\footnote{Details on the datasets, target systems, and hardware are given in Appendix~\ref{app:detailedsetting}.} in Table~\ref{tab:api} and get the final transcription in return according to our proposed algorithms. 
To access the commercial cloud APIs, we first register on the platforms and then use audio AEs to query the speech-to-text services according to API keys provided by the platforms.
For the attacks against SR services, we conduct our targeted attack on three systems (also in Table~\ref{tab:api}). 
For Microsoft SI and Jingdong SV, we choose ten people from Voxceleb dataset~\cite{DBLP:conf/interspeech/NagraniCZ17} and enroll four utterances for each person.
Since Microsoft SV is text-dependent, we collect five volunteers' voice data, and each volunteer is required to enroll 10 fixed sentences.
To generate AEs, we query the APIs for 10,000 times for each target person.

\bluea{
To evaluate the performance of NI-Occam,~\ie, our non-interactive physical attacks on voice control devices, 
we design two sets of experiments: attacks on popular commercial voice assistants and human perception of the audio AEs. 
We evaluate NI-Occam on Google Assistant, Apple Siri, Microsoft Cortana, iFlytek, and Amazon Echo, as shown in Table~\ref{tab:VCD}. We generate 10 sets of audio AEs locally, whose target phrases of AEs are the same as those in the evaluation of Occam. We then play them using a JBL Clip 3 portable speaker near the target devices in a quite laboratory. The distance between the speaker and the target devices is around 15cm.}

\begin{table}[t]\small
%\renewcommand\arraystretch{1.2}
% \centering
\caption{Details of the commercial speech and speaker recognition services.}
\vspace{-3mm}
\begin{tabular}{c|c|c}
\hlinew{1.2pt}
\textbf{Task }                                                                  & \textbf{ Commercial Services }  & \textbf{ Return results} \\
\hlinew{1.2pt}
\multirow{5}{*}{\begin{tabular}[c]{@{}c@{}}Speech \\Recognition\\\end{tabular}} & Google STT$^\ddag$~\cite{Google}            & D$^\flat$+S$^\natural$    \\
\cline{2-3}
                                                                                & Microsoft ASR~\cite{Microsoft}          & D+S                                                            \\
%\cline{2-3}
%                                                                                & IBM STT~\cite{IBM}                  & D+S                                                            \\
\cline{2-3}
                                                                                & Alibaba SSR$^\S$~\cite{Alibaba}             & D                                                                 \\
\cline{2-3}
                                                                                & Tencent SSR~\cite{Tencent}              & D                                                                 \\
\cline{2-3}
                                                                                & iFlytek~\cite{iFlytek}                  & D                                                               \\
\hlinew{1.2pt}
\multirow{3}{*}{\begin{tabular}[c]{@{}c@{}}Speaker \\Recognition \end{tabular}} & Microsoft SI$^\sharp$~\cite{Microsoft}      & D+S                                    \\
\cline{2-3}
                                                                                & Microsoft SV$^\dag$~\cite{Microsoft}       & D+S                                   \\
\cline{2-3}
                                                                                & Jingdong SV~\cite{Jingdong}              & D                                       \\
\hlinew{1.2pt}
\end{tabular}\label{tab:api}

\begin{tablenotes}
\item\footnotesize{
Note that, (i) $\ddag$: ``STT'' means ``Speech-to-Text''. We use the command\_and\_search model in Google STT API. (ii) $\S$: ``SSR" means ``Short Speech Recognition''. (iii) $\sharp$: ``SI'' means ``Speaker Identification''. (iv) $\dag$: ``SV'' means ``Speaker Verification''. (v) $\flat$: ``D'' means the model returns the final decision. (vi) $\natural$: ``S'' means the model returns the confidence score or confidence level.
}
  \end{tablenotes}
    % \footnotesize{Note that, (i) $\ddag$: ``STT'' means ``Speech-to-Text''. We use the command\_and\_search model in Google STT API. (ii) $\S$: ``SSR" means ``Short Speech Recognition''. (iii) $\sharp$: ``SI'' means ``Speaker Identification''. (iv) $\dag$: ``SV'' means ``Speaker Verification''. (v) $\flat$: ``D'' means the model returns the final decision. (vi) $\natural$: ``S'' means the model returns the confidence score or confidence level.}

%Although some APIs provide confidence scores, our attack does not require such information.}

\vspace{-3mm}

\end{table}

\begin{table}[t]\small
\centering
\caption{Details of the commercial voice control devices.}
\vspace{-3mm}
\begin{tabular}{c|c|c|c|c} 
\hlinew{1.2pt}
\begin{tabular}[c]{@{}c@{}}\textbf{Voice}\\\textbf{Assistant}\end{tabular} & \textbf{Version} & \textbf{Device}                                      & \begin{tabular}[c]{@{}c@{}}\textbf{Audio}\\\textbf{Source}\end{tabular}                                    & \textbf{Speaker}            \\ 
\hlinew{1.2pt}
Apple Siri                                                                 & 13.6.1            & iPhone 11                                            & \multirow{8}{*}{\begin{tabular}[c]{@{}c@{}}Default \\media \\player,\\ThinkPad \\X1 Carbon\end{tabular}} & \multirow{8}{*}{JBL Clip3}  \\ 
\cline{1-3}
iFlytek                                                                    & 10.0.8            & iPhone 11                                            &                                                                                                            &                             \\ 
\cline{1-3}
\begin{tabular}[c]{@{}c@{}}Microsoft\\Cortana\end{tabular}                & 3.3.3             & \begin{tabular}[c]{@{}c@{}}Samsung\\C9\end{tabular}  &                                                                                                            &                             \\ 
\cline{1-3}
\begin{tabular}[c]{@{}c@{}}Google\\Assistant\end{tabular}                 & 2.5.1             & \begin{tabular}[c]{@{}c@{}}Nokia\\7plus\end{tabular} &                                                                                                            &                             \\ 
\cline{1-3}
\begin{tabular}[c]{@{}c@{}}Amazon\\Echo\end{tabular}                                                                      &    631499520   & \begin{tabular}[c]{@{}c@{}}Echo\\1st gen\end{tabular} &                                                                                                            &                             \\
\hlinew{1.2pt}
\end{tabular}\label{tab:VCD}
\end{table}

%\textbf{Word Error Rate (WER).} We use the word error rate to evaluate the effectiveness of the proposed attack. WER calculates the difference between the original word sequence and the transcribed word sequence. Given the number of inserted words $I$, the number of substitute words $S$, the number of deleted words $D$, and the total number of words in the sentence $N$, WER is computed as
%\begin{equation}
%  WER=\frac{I+S+D}{N}.
%\end{equation}
%\textbf{Success Rate of Attack (SRoA).}
\noindent \textbf{Methods for Comparison.}
We compare Occam with two state-of-the-art black-box adversarial attacks against ASR and SR services,~\ie,~% \reda{Genetic Algorithm-based Attack (GAA)~\cite{Taori}, Selective Gradient Estimation Attack (SGEA)}~\cite{TIFS-SGEA},
Devil's Whisper~\cite{Whisper} and FakeBob~\cite{DBLP:journals/corr/abs-1911-01840}. 
%Since GAA and SGEA cannot generate AEs without the logits (\ie, prediction scores), we only evaluate them on DeepSpeech.
Since Devil's Whisper originally utilized confidence scores to filter the \reda{synthetic audio data}, in the follow-up evaluations, we omit this step to adapt Devil's Whisper to the decision-based attack.
%since quite a lot commercial services only return the final decisions.
To better evaluate the effectiveness of Occam, we also select five decision-based attacks % with the same threat scenario 
for comparison: the boundary attack~\cite{BoundaryAttack}, the HopSkipJump attack (HSJA)~\cite{HopSkipJumpAttack}, the opt-attack~\cite{query_efficient}, the evolutionary attack\footnote{In our experiments, the bilinear interpolation in the evolutionary attack is removed in the audio domain as it is not applicable to the time-series speech data.}~\cite{CVPR}, and the Differential Evolution attack (DEA)\footnote{We give a detailed description on DEA in Appendix~\ref{app:DEA}.}~\cite{price2013differential}. %Note that, HSJA is an improved version of the boundary attack, and the evolutionary attack is constructed based on CMA-ES. In our experiments, the bilinear interpolation in the evolutionary attack is removed in the audio domain as bilinear interpolation is not applicable to the time-series speech data. 

\bluea{
We compare NI-Occam with a straightforward non-interactive attack,~\ie, the superposition attack. %\footnote{Note that, NI-Occam is the first effective non-interactive physical attack in the literature.}. 
In this attack, we just directly superimpose the original example and the command audio. Since this procedure does not require any knowledge of the target systems, the superposition attack is non-interactive. 
}

\noindent\textbf{Evaluation Metrics.}
%The metrics used in our experiments are presented in Table~\ref{tab:metric}.
We use the success rate of attack (SRoA) to evaluate the effectiveness of AEs. SRoA calculates the proportion of AEs that can successfully attack ASR or SR services. %Considering that the goals of successfully attacking the ASR and SR systems are different, we further define three sub-SRoAs,~\ie, SRoA-ASR, SRoA-SI, and SRoA-SV.
Besides, we use SNR\footnote{SNR calculates the ratio of the signal power of the original audio to the noise power. A higher SNR indicates a lower noise level.} to describe the perturbation on audio AEs and the number of queries on the target model to indicate the efficiency of the attacks. 
It is worth noting that the success rate or the SNR is not the only metric that determines whether an AE can successfully attack the target system. An effective AE should fool both the model and the human. Hence, we should combine both the SRoA and SNR when evaluating the effectiveness of AEs. 
\bluej{Note that, to illustrate this statement, we conduct a user study to analyze how SNRs affect the human perception. The results show that a lower SNR makes it easier for the users to notice or even recognize the AEs. Due to the space limit, we present the detailed results in Appendix~\ref{app:SNR-human}.}

\subsection{Evaluation on Cloud Speech APIs}\label{sec:api}

\begin{table*}[t]\small
%\renewcommand\arraystretch{1.2}
% \centering
\caption{Experimental results on targeted attacks on commercial cloud speech-to-text APIs.}
\vspace{-3mm}
\begin{tabular}{c|cc|cc|cc|cc|cc|cc|cc}
\hlinew{1.2pt}
\multirow{2}{*}{\textbf{\begin{tabular}[c]{@{}c@{}}Cloud service\end{tabular}}} & \multicolumn{2}{c|}{\textbf{\begin{tabular}[c]{@{}c@{}}Boundary \\ Attack\end{tabular}}} & \multicolumn{2}{c|}{\textbf{Opt-Attack}} & \multicolumn{2}{c|}{\textbf{\begin{tabular}[c]{@{}c@{}}Evolutionary \\ Attack\end{tabular}}} & \multicolumn{2}{c|}{\textbf{HSJA}} & \multicolumn{2}{c|}{\textbf{DEA}} & \multicolumn{2}{c|}{\textbf{\begin{tabular}[c]{@{}c@{}}Devil's \\ Whisper\end{tabular}}} & \multicolumn{2}{c}{\textbf{Occam}} \\ \cline{2-15} 
                                                                                & SRoA                                        & SNR                                        & SRoA                & SNR                & SRoA                                          & SNR                                          & SRoA             & SNR             & SRoA            & SNR             & SRoA                                        & SNR                                        & SRoA        & SNR                        \\ \hlinew{1.2pt}
\textbf{Google STT}                                                                 & 10/10                                       & 6.80                                        & 10/10               & 5.52               & 10/10                                         & 6.11                                         & 10/10            & 5.38            & 10/10           & 6.52            & 5/10                                        &   {9.13}                                        & \textbf{10/10}       & {{\textbf{15.33}}}       \\ \hline
\textbf{Microsoft ASR}                                                              & 10/10                                       & 8.71                                       & 10/10               & 6.45               & 10/10                                         & 6.60                                          & 10/10            & 4.59            & 10/10           & 7.35            & 8/10                                        & {9.91}                                       & \textbf{10/10}       & {\textbf{10.74}}                      \\ \hline
%\textbf{IBM STT}                                                                    & 10/10                                       & 0.55                                       & 10/10               & -1.64              & 10/10                                         & -1.11                                        & 10/10            & 0.47            & 10/10           & -1.61           & 2/10                                        &  \underline{\textbf{7.42}}                                        & 10/10       &  \textbf{5.07}                  \\ \hline
\textbf{Alibaba SSR}                                                            & 10/10                                       & {9.32}     & 10/10               & 5.70                & 10/10                                         & 7.20                                          & 10/10            & 8.11            & 10/10           & 5.44            & 3/10                                        & 8.21                                       & \textbf{10/10}       & {\textbf{17.84}}                   \\ \hline
\textbf{Tencent SSR}                                                            & 10/10                                       & {11.03}                                      & 10/10               & 6.43               & 10/10                                         & 9.14                                         & 10/10            & 8.49            & 10/10           & 6.06            & 4/10                                        & 9.64                                       & \textbf{10/10}       & {\textbf{16.70}}                        \\ \hline
\textbf{iFlytek}                                                                & 10/10                                       & 6.78                                       & 10/10               & 6.73               & 10/10                                         & 6.50                                          & 10/10            & 5.08            & 10/10           & 6.47            & 7/10                                        & {9.27}                                       & \textbf{10/10}       & {\textbf{11.22}}                    \\
\hlinew{1.2pt}
\textbf{Average}      & 10/10                                       & 8.53                                       & 10/10               & 6.17               & 10/10                                         & 7.11                                          & 10/10            & 6.33            & 10/10           & 6.37            & 5.4/10                                        & {9.23}                                       & \textbf{10/10}       & {\textbf{14.37}}                    \\
\hlinew{1.2pt}
\end{tabular}\label{tab:apisnr}

\begin{tablenotes}
\item    \footnotesize{
Note that, 
(i) Devil's Whisper originally utilized confidence scores to filter the synthetic audio data. Since the confidence scores are not available from Alibaba, iFlytek, and Tencent, we have omitted the score-related processing step in reproduced experiments of Devil’s Whisper on these three speech APIs. Targeting Google TTS and Microsoft ASR (which return confidence scores), we follow the original version of Devil's Whisper when conducting the experiments on Google and Microsoft.
(ii) In reproduced experiments of Devil's Whisper, we have used 14,569 (filtered) clips, containing different commands as those adopted in Devils Whisper paper. Moreover, we have used 3,000 extra clips for certain commands with unsatisfactory clip quality. The total length of the TTS audio corpus and the supplemental corpus of the Mini LibriSpeech dataset are 6.8 hours and 7.3 hours, respectively. 
%Devil's Whisper originally utilized confidence scores to filter the synthetic audio data. We omit this step to adapt Devil's Whisper to the decision-based attack.
 }
  \end{tablenotes}

\vspace{-3mm}
\end{table*}
%In our experiment, the WER for untargeted attacks needs to be higher than 3 to ensure that audio AEs are significantly misrecognized.[HOW?]

%Fig.~\ref{fig:spec} shows the waveforms and spectrograms of the original audio and the adversarial audios from the four attacks.
%We can see that the waveforms of the original audio and the audio AE from our attack are almost the same. However, the differences in other attacks are more noticeable and therefore more likely to be perceived by humans. 

% \begin{figure}[t!]
%   \centering
%   \includegraphics[width=0.95\linewidth]{waveformcomparsion_short.pdf} \\
%   \vspace{-2mm}
%   \caption{\bluea{Waveforms and spectrograms of the original audio and adversarial audios generated by the targeted attacks against Alibaba SSR. The audio AEs can be recognized as ``call my wife''. }}
%   \vspace{-5mm}
%   \label{fig:spec}
% \end{figure}

\begin{table*}[ht]\small
% \centering
%\renewcommand\arraystretch{1.2}
\caption{Experimental results on targeted and untargeted attacks on cloud speaker recognition APIs.}
\vspace{-3mm}

\begin{tabular}{c|cc|cc|cc|cc|cc|cc|cc}
\hlinew{1.2pt}
\multirow{2}{*}{\textbf{Cloud service}}                                               & \multicolumn{2}{c|}{\begin{tabular}[c]{@{}c@{}}\textbf{Boundary}\\\textbf{Attack}\end{tabular}} & \multicolumn{2}{c|}{\textbf{Opt-Attack}} & \multicolumn{2}{c|}{\begin{tabular}[c]{@{}c@{}}\textbf{Evolutionary }\\\textbf{Attack}\end{tabular}} & \multicolumn{2}{c|}{\textbf{HSJA}} & \multicolumn{2}{c|}{\textbf{DEA}} & \multicolumn{2}{c|}{\textbf{FakeBob}} & \multicolumn{2}{c}{\textbf{Occam}}  \\ 
\cline{2-15}
                                                                                    & SRoA  & SNR                                                                                      & SRoA  & SNR                              & SRoA  & SNR                        & SRoA  & SNR                                                                                          & SRoA  & SNR                       & SRoA  & SNR                           & SRoA  & SNR                               \\ 
\hlinew{1.2pt}
\textbf{Microsoft SI}                                                               & 10/10 & 7.13                                                                                     & 10/10 & 7.01                             & 10/10 & {7.25}                       & 10/10 & 4.87                                                                                         & 10/10 & 2.84                      & 0/200 & N/A                           & \textbf{10/10} & {\textbf{14.31}}                             \\ 
\hline
\textbf{Microsoft SV}                                                               & 10/10 & 8.48                                                                                     & 10/10 & 5.85                             & 10/10 & {9.93}                       & 10/10 &4.27                                                                                        & 10/10 & 3.05                      & 0/200 & N/A                           & \textbf{10/10} & {\textbf{13.25}}                              \\ 
\hline
\textbf{Jingdong SV}                                                                & 10/10 & 6.26                                                                                     & 10/10 & 6.42                             & 10/10 & 5.97                       & 10/10 & 5.89                                                                                         & 10/10 &     5.67                      & 2/200 & {{35.30}}                      & \textbf{10/10} & \textbf{13.78}                             \\ 
\hline
\begin{tabular}[c]{@{}c@{}}\textbf{Microsoft SI~}\\\textbf{(untarget)}\end{tabular} & 10/10 & 9.61                                                                                     & 10/10 &9.95                          & 10/10 & 9.04                       & 10/10 & 9.43                                                                                         & 10/10 & 6.53                      & 2/200   & {{40.21}}                            & \textbf{10/10} & \textbf{14.92}                           \\
\hlinew{1.2pt}
\textbf{Average}      & 10/10                                       & 7.87                                       & 10/10               & 7.31               & 10/10                                         & 8.05                                        & 10/10            & 6.12            & 10/10           & 4.52            & 1/200                                       &  {{37.76}}                                      & \textbf{10/10}       & \textbf{14.07}                    \\
\hlinew{1.2pt}
\end{tabular}
\label{tab:sr}

\begin{tablenotes}
\item    \footnotesize{
Note that, 
(i) we only evaluate untargeted attacks against Microsoft Azure SI. For speaker verification systems, the goal of untargeted attacks is the same as that of targeted attacks.
(ii) N/A denotes ``not available''. Since there is no effective AE against Microsoft SI and SV in FakeBob, the SNR is not available.

}
  \end{tablenotes}

% \footnotesize{Note that, 
% (1) We only evaluate untargeted attacks against Microsoft Azure SI. For speaker verification systems, the goal of untargeted attacks is the same as that of targeted attacks.

% (2) N/A denotes ``not available''. Since there is no effective AE against Microsoft SI and SV in FakeBob, the SNR is not available.

% }

\end{table*}

\noindent\textbf{Effectiveness of Occam on ASR services.}
Table~\ref{tab:apisnr} shows the performance of the targeted attacks on different commercial speech services after 30,000 queries \reda{(10,000 queries on Google)}.
%\reda{The attacks on the commercial APIs require fewer queries to generate an audio AE with small noises than that on DeepSpeech. This may be because commercial APIs will automatically check the spelling of the generated texts while DeepSpeech won't do it.}
\bluea{With regard to SNR, Occam performs the best among the seven attacks, reaching the best SNR of 17.84dB. Notably, DEA only obtains an average SNR of 6.37dB, which means that the perturbations of the audio AEs are very large. 
The results demonstrate that as the gradient-free optimization method in Occam's cooperative co-evolution framework, CMA-ES is a better choice than the differential evolution. The reason behind this may be that CMA-ES is more suitable than DEA to solve the non-separable optimization, since CMA can well learn the dependencies between variables. 
However, the average SNR of AEs generated by the evolutionary attack (which is based solely on CMA-ES) can only achieve 7.11dB, while Occam has an average SNR of 14.37dB. 
This indicates that although CMA-ES can solve the non-separable optimization problem, when regarding to complex speech data, CMA-ES becomes ineffective to solve the discontinuous large-scale global optimization problem.
Hence, CMA-ES alone cannot deal with complex speech data well. 
The above results demonstrate that Occam can effectively manage high-dimensional audio data. 
%We notice that the SNRs on IBM STT are relatively low. This is because IBM STT itself cannot accurately recognize normal audios,~\ie, only about 25\% accuracy. Thus, the AEs against IBM STT cannot perform well, either. 
%Another interesting observation is that HSJA, as an improved version of the boundary attack, performs much worse than the boundary attack. Unlike the boundary attack that uses random guessing to locate the adversarial area after the binary search, HSJA instead uses the binary information at the decision boundary to estimate the gradient of the right direction after the binary search. However, in (A)SR systems, the exact matching of the transcript lowers the accuracy of the sampling estimation, making the binary information less accurate for HSJA. Hence, HSJA is less efficient than the boundary attack in generating audio AEs.
}

Compared to Devil's Whisper, Occam can achieve 100\% SRoAs on all API services, while Devil's Whisper can only achieve an average SRoA of 54\%. 
\bluea{This indicates that Devil's Whisper is less effective in fooling the target speech recognition model than Occam.}
Figure~\ref{fig:specfull} in Appendix~\ref{app:wave} shows the waveforms and spectrograms of the original audio and the adversarial audios generated from Occam and Devil's Whisper.
% \footnote{Due to space limitation, this figure only shows the waveforms of Occam and Devil's Whisper. A full version of the comparisons with other attacks is given in Appendix~\ref{app:wave}.}. 
We can see from the waveforms that the audio AE generated from Occam is almost the same as the original audio. However, the differences in Devil's Whisper are more noticeable and thus more likely to be perceived by humans. 
Besides, %Devil's Whisper is a targeted attack and cannot be directly transformed into an untargeted attack.
although we only choose ten commands in the experiments, Occam can generate AEs of arbitrary phrases\footnote{To illustrate this, we further evaluate Occam on a large group of target phrases. Related results are presented in Appendix~\ref{app:large}.}, while Devil's Whisper can only generate a limited number of target phrases with a trained model.
These observations suggest our decision-based Occam is more effective, powerful, and practical. 
We also evaluate the untargeted attacks. Due to space limitation, related results are given in Appendix~\ref{app:wave}.

\begin{table}[t]\small
\centering
\caption{\bluea{Experimental results of the physical attacks.}}\label{tab:physicalattackavg}
\vspace{-4mm}
\begin{tabular}{c|c|c|c|c|c|c} 
\hlinew{1.2pt}
\multicolumn{2}{c|}{\textbf{Devices}}                                                                                                                             & \textbf{Siri} & \textbf{iFlytek} & \textbf{Cortana} & \textbf{Google} & \textbf{Echo}  \\ 
\hlinew{1.2pt}
\multirow{3}{*}{\textbf{NI-Occam}}                                                                            & SRoA                                               & 6/10          & 6/10             & 6/10             & 4/10            & 4/10           \\ 
\cline{2-7}
                                                                                                              & \begin{tabular}[c]{@{}c@{}}SNR\\(dB)\end{tabular}  & 9.81          & 9.09             & 9.58             & 10.36           & 9.42           \\ 
\hline
\multirow{2}{*}{\begin{tabular}[c]{@{}c@{}}\textbf{Super-}\\\textbf{position}\\\textbf{~attack}\end{tabular}} & SRoA                                               & 5/10          & 1/10             & 1/10             & 1/10            & 2/10           \\ 
\cline{2-7}
                                                                                                              & \begin{tabular}[c]{@{}c@{}}SNR \\(dB)\end{tabular} & 7.00          & 7.00             & 7.00             & 7.00            & 7.00           \\
\hlinew{1.2pt}
\end{tabular}
\end{table}
%\subsubsection{Effectiveness of Occam on SR systems}

\noindent \textbf{Effectiveness of Occam on SR services.}
%\subsection{Evaluation on Attacks against Speaker Recognition Systems}\label{sec:ExperimentSR}
We evaluate the SNRs and SRoAs of the AEs against SR services, as in Table~\ref{tab:sr}. 
Among the attacks against SR services, Occam still achieves 100\% SRoAs, indicating that the audio AEs can be successfully recognized as the target person (or mislead the SR system). FakeBob can only achieve a 1\% SRoA. Note that we tested FakeBob with 200 instead of 10 AEs. Since the success rate of FakeBob is too low, we enlarge the AE set to create an effective AE. We find that the results given by FakeBob are higher than those in Table~\ref{tab:sr}. This is probably because the upgrading of Microsoft speaker recognition systems makes the transferability of the AEs in FakeBob no longer effective. As for SNR, the AEs from Occam can achieve pretty good SNRs, outperforming other decision-based attacks. %The results show that the perturbations of the AEs from our attack are more subtle than those from the other five attacks.
Although effective AEs from FakeBob can achieve SNR as high as 35.3dB, the SRoA is too low to effectively mislead the SR services and thus far from being practical.

\subsection{Evaluation of NI-Occam against Voice Control Devices}\label{sec:evaluation-NI}

\begin{table*}[ht]\small
% \centering
\caption{\bluea{Evaluation results on human perception of the non-interactive physical attacks.}}\label{tab:physicalhuman}
\vspace{-3mm}
\begin{tabular}{c|c|c|c|c|c|c} 
\hlinew{1.2pt}
\textbf{Devices}                              & \textbf{Method}      & \textbf{Normal (\%)} & \textbf{Noise (\%)} & \textbf{Talking (\%)} & \textbf{Once-recognize (\%)} & \textbf{Twice-recognize (\%)}  \\ 
\hlinew{1.2pt}
\multirow{2}{*}{\textbf{Apple Siri}}         & NI-Occam             & 12.4                 & 61.7                & 25.9                  & 4.1                          & 5.9                            \\ 
\cline{2-7}
                                              & Superposition attack & 0.0                  & 0.0                 & 100                   & 88.1                         & 91.9                           \\ 
\hlinew{1.2pt}
\multirow{2}{*}{\textbf{iFlytek}}            & NI-Occam             & 9.6                  & 58.3                & 32.1                  & 5.1                          & 6.9                            \\ 
\cline{2-7}
                                              & Superposition attack & 0.0                  & 0.0                 & 100                   & 86.5                         & 94.6                           \\ 
\hlinew{1.2pt}
\multirow{2}{*}{\textbf{Mircrosoft Cortana}} & NI-Occam             & 14.2                 & 56.7                & 29.1                  & 6.1                          & 8.3                            \\ 
\cline{2-7}
                                              & Superposition attack & 0.0                  & 0.0                 & 100                   & 83.8                         & 86.5                           \\ 
\hlinew{1.2pt}
\multirow{2}{*}{\textbf{Google Assistant}}   & NI-Occam             & 15.0                 & 55.7                & 29.4                  & 4.8                          & 8.2                            \\ 
\cline{2-7}
                                              & Superposition attack & 0.0                  & 0.0                 & 100                   & 83.8                         & 86.5                           \\ 
\hlinew{1.2pt}
\multirow{2}{*}{\textbf{Amazon Echo}}        & NI-Occam             & 2.7                  & 68.2                & 29.1                  & 12.2                         & 14.2                           \\ 
\cline{2-7}
                                              & Superposition attack & 0.0                  & 0.0                 & 100                   & 86.5                         & 94.6                           \\
\hlinew{1.2pt}
\end{tabular}

\

\begin{tablenotes}
\item    \footnotesize{
Note that, 
(i) ``Normal'' means that the volunteer regards the audio as a normal audio. 
(ii) ``Noise'' means that the volunteer can feel some noises. 
(iii) ``Talking'' means that the volunteer can hear talking in the audio. If the volunteer thinks there is talking in the audio, he/she is then asked to recognize the content of the talking. 
(iv) The audio will be labeled as ``once-recognize'' or ``twice-recognize'' if the volunteer recognizes over half of the content after listening to the audio once or twice, respectively.
 }
  \end{tablenotes}

%     \footnotesize{
% Note that, 
% ``Normal'' means that the volunteer regards the audio as a normal audio. 
% ``Noise'' means that the volunteer can feel some noises. 
% ``Talking'' means that the volunteer can hear talking in the audio. If the volunteer thinks there is talking in the audio, he/she is then asked to recognize the content of the talking. 
% This audio will be labeled as ``once-recognize'' or ``twice-recognize'' if the volunteer recognizes over half of the content after listening to the audio once or twice, respectively.}
\vspace{-3mm}
\end{table*}

\noindent \textbf{Effectiveness. }
We test the effectiveness of NI-Occam on 5 voice control devices, and the results are given in Table~\ref{tab:physicalattackavg}\footnote{\bluej{Detailed results on individual commands can be found in Appendix \ref{app:attempts} (see Table~\ref{tab:physicalattack}).}}. 
If the audio AE can be correctly recognized by the devices as the target command within 3 attempts (play the AEs within three times), we consider this AE successful. 
Overall, NI-Occam achieves an average SRoA of 52\% and SNR of 9.65dB.  
Note that NI-Occam is a non-interactive attack that requires no access to the target devices, which is very practical since some devices like Apple Siri do not provide a programmable API. For example, Devil's Whisper~\cite{Whisper} fails to generate effective AEs when confronted with devices that do not return confidence scores. We also find that NI-Occam performs well on Apple Siri, while Devil's Whisper failed in attacking updated versions of Siri. This indicates that NI-Occam is more effective in leveraging the useful transferability of the AEs and can generate successful AEs with minimal information from the target devices.
We also evaluate a simple non-interactive attack as the baseline,~\ie, the superposition attack. Since the AEs from the superposition attack are constructed by superimposing two audios, the SNR is adjustable. We set the SNRs as 7.00dB, smaller than those of NI-Occam. A smaller SNR means the audio of the target command is more obvious in the audio AE, making it easier for the devices to recognize. However, the superposition attack can only achieve an average SRoA of 20\% with a 7.00dB SNR. More importantly, the superposition attack is easily perceived by human, which we will discuss shortly.

%We notice that the physical-effective AEs fail to attack the commercial services in the digitial domain. This interesting finding shows the limitation of audio AE's transferability,~\ie, the AEs become ineffective if the target model is not similar enough to the local model. Therefore, here we propose a potential research direction to improve the adaptability of the physical attack. Considering that the AEs from our co-evolutionary attack is generated by crossing the decision boundary, the boundary information is useful for the local model in the transfer-based attack to approximate the target black-box systems. In turn, the gradient information obtained from the local model can also help the co-evolutionary attack find better search directions. Thus, integrating the co-evolutionary attack with the transfer-based attack may help make the AEs more robust and effective in the physical domain. We believe this is a promising direction to enhance our proposed physical attack, and will leave it as our future work. 

\bluej{\qi{We also evaluate the impact of the number of attempts on SRoA with a larger group of target phrases. We observe that increasing the query attempts can help increase SRoA to 70\%, and our NI-Occam can also perform well on large sets of 60 commands with an SRoA of 71.7\%. Detailed results are given in Appendixes~\ref{app:attempts} and~\ref{app:large}.
}}

\noindent\textbf{Human Perception. }\label{sec:human}
\bluea{
Although SNR describes the proportion of noise, it cannot fully reflect the imperceptibility of the audio. For example, if the noise can well fit the background, even though the signal has a low SNR, the users cannot perceive it. On the contrary, if the noises all appear in a small piece of audio, although the overall SNR is high, the users may readily perceive the command.}

\bluea{
To evaluate the performance of NI-Occam on human perception, we surveyed 37 volunteers aged from 19 to 24 (who are sensitive to sound), including 21 males and 16 females. In the experiment, we first show some examples of ``noise'', ``normal'', ``talking'', and ``recognized'', and then ask the volunteers to listen to the audio AEs and tell their views about them. }
Specifically, 
%the volunteers need to listen to several pieces of audio AEs generated from NI-Occam and the superposition attack. 
the audio AEs are generated from NI-Occam and the superposition attack and can successfully attack the devices. 
Each volunteer ranks 6, 6, 6, 4, and 4 successful AEs from NI-Occam on Apple Siri, iFlytek, Microsoft Cortana, Google Assistant, and Amazon Echo, respectively\footnote{The AEs are the ones that can fool the devices, as shown in Table~\ref{tab:physicalattackavg}.}. %As for the superposition attack, each volunteer ranks 5, 1, 1, 1, and 2 successful AEs on Apple Siri, iFlytek, Microsoft Cortana, Google Assistant, and Amazon Echo, respectively.

\bluea{
Table~\ref{tab:physicalhuman} presents the results of the experiments on human perception. 
Overall, NI-Occam performs much better than the superposition attack.
More than 67\% volunteers think the audio generated from NI-Occam is normal or just noisy, while 100\% volunteers can recognize the AEs from the superposition attack. 
This is because the audios crafted in the superimposing manner can be more easily noticed due to the human ear's excellent ability of speech separation. 
We also assume that a proper noise level is acceptable because the background environment may be noisy, and the equipment may emit some noise due to a temporary fault. The results show that NI-Occam is stealthy enough and cannot be easily perceived.}

\section{Related Work}\label{sec:relatedwork}

\subsection{Audio Adversarial Examples}\label{sec:aaasrs}
\noindent \textbf{Adversarial Examples Against ASR Systems.} Despite the great success of adversarial examples in the image domain, the transcription of spontaneous speech poses a more significant challenge for crafting audio AEs. The early results have clearly indicated that ASR systems are inherently vulnerable to AEs in white-box settings. Among others, Carlini {\em et al.}~\cite{CWaudio} was the first to implement an iterative optimization-based attack with a success rate of $100\%$ on the end-to-end Mozilla DeepSpeech model. However, it takes approximately one hour for their attack to produce one adversarial example on a single NVIDIA 1080Ti GPU, and the crafted adversarial sample fails when being played over the air. Concurrently, CommanderSong~\cite{CommanderSong}, which embedded malicious commands into popular songs, was reported to have successfully attacked against Kaldi~\cite{Kaldi}. It further launched a very limited over-the-air attack, which is heavily dependent on the recording devices, speakers and room settings. Efforts were also made by~\cite{DBLP:conf/ijcai/YakuraS19} to build robust over-the-air AEs by utilizing impulse responses to simulate the reverberation with a success rate of around 60\%. \redj{Moreover, Chen {\em et al.}~\cite{chenMetamorphNDSS}
achieved a 90\% success rate of over-the-air attacks over the attack distance of up to 6m by further removing device- and environment-specific features.
} Besides, imperceptible audio AEs were produced in~\cite{schonherr2018adversarial} via the psychoacoustic model. Both imperceptible and robust audio AEs were constructed in~\cite{imperceptible} against the Lingvo ASR system, with a success rate of 50\%. 

%{\color{blue} To address the time  Liu {\em et al.}~\cite{liu2019adversarial} introduced weighted-sampling perturbation to reduce the time cost of constructing audio AEs~\cite{CWaudio} from hours to minitues.}

Compared to the above attacks, our attacks have the following superiorities: 1) Our attacks are more practical since these attacks have access to the internal information of target model; 2) Instead of only targeting one open-source system in the white-box setting, our attacks evaluate the robustness of many representative audio processing systems in real-world scenarios.
%3) Our attack is generic and effective for both commercial ASR and SR systems.

While adversarial example based white-box attacks have obtained excellent results against open-source ASR systems, its black-box counterpart hasn't \reda{made} big progress until recently. By leveraging the last layer (\textit{i.e.}, logits) of DNNs inside the DeepSpeech, Taori {\em et al.}~\cite{Taori} obtained the fitness score of adversarial inputs, and combined genetic algorithms with the gradient estimation to attack DeepSpeech. In addition to the rather low success rates even after 300,000 queries, this type of black-box attack is not applicable to commercial systems. \reda{Following this work, selective gradient estimation attack \cite{TIFS-SGEA} was proposed to achieve a success rate of 98\% in this setting.}  Moreover, multi-objective genetic algorithms were also introduced to start a black-box adversarial attack against ASR systems~\cite{KAM2018}. However, due to the relatively large word error rate (WER) after many evolutions in the attack, it becomes ineffective for generating audio AEs. A very recent work, Devil{\textquoteright}s Whisper~\cite{Whisper}, utilized confidence scores exposed by commercial ASR systems to launch the black-box attack. However, it is worth noting that there are many commercial ASR systems that do not return any score information, and service providers are also apt to hide these scores to reduce the risk of adversarial attacks, considering that these information almost has no effect on the user experience and may be mainly exploited by malicious attackers.

There are two key differences: 1) Our Occam, as a generic attack, requires only the final decisions to generate audio AEs, effective to both commercial ASR and SR services; 2) Our NI-Occam requires no access to the targeted devices, but still can generate effective audio AEs. Our attacks are more practical in real world scenarios.

\noindent \textbf{Adversarial Examples Against SR Systems.} When it comes to adversarial example generation against SR systems, relatively little work has been done in both white-box and black-box cases. Kreuk {\em et al.}~\cite{DBLP:conf/icassp/KreukACK18} presented a white-box adversarial attack against the DNN-based speaker verification system. Gong {\em et al.}~\cite{DBLP:journals/corr/abs-1711-03280} demonstrated the vulnerability of speaker identification system to audio AEs in the white-box scenario. Obviously, these attacks require access to internal structures and parameters of the target systems, and they are impractical when facing commercial SR systems.

In a recent work, SirenAttack~\cite{du2019sirenattack} was presented to launch a black-box attack against a number of classification-oriented acoustic systems, including the SR system via the predicted probabilities/scores. More recently, FakeBob~\cite{DBLP:journals/corr/abs-1911-01840} also utilized the predicted probabilities/scores to conduct a black-box adversarial attacks against Talentedsoft API~\cite{Talentedsoft} with a success rate of 100\%. The main limitation of SirenAttack and FakeBob is that they lose the effectiveness when applied to commercial SR systems, which usually hide the predicted scores. For example, the Microsoft Azure SR API service only provides the decision (\textit{i.e.}, the predicted speaker) along with three confidence levels (\textit{i.e.}, low, normal, or high) to users. Our attack shows its superiority to prior attacks by achieving an attack success rate of 100\% against commercial SR systems even if the service provider conceals the prediction score information.

\vspace{-10pt}

\subsection{Other Types of Attacks}\label{sec:othertypes}
In addition to audio AEs, researchers have also discovered that intelligent voice systems are vulnerable to other types of attacks, including misinterpretation attack and hidden voice attack.

\noindent \textbf{Misinterpretation Attacks.}
Kumar {\em et al.}~\cite{DBLP:conf/uss/KumarPMHMBB18} conducted an empirical analysis of misinterpretations and investigated security implications on Amazon Alexa, based on which they introduced a new attack called skill squatting to surreptitiously route users to malicious third-party public services. Along this direction, Zhang {\em et al.}~\cite{zhang2019dangerous} reported similar attacks against ASR systems, where a malicious skill with the similarly pronounced name or paraphrased name was exploited to impersonate a benign skill. Also targeting at ASR systems, Zhang {\em et al.}~\cite{DBLP:conf/ndss/ZhangXMYCG19} designed a linguistic-model guided fuzzing tool called LipFuzzer to systematically discover misinterpretations leading to malicious attacks.

\noindent \textbf{Hidden Voice Attacks}.
Besides, by either exploiting knowledge of the feature extraction algorithm or hardware vulnerabilities in microphone circuits, the adversary can embed hidden commands into an audio carrier, in the form of noises, thereby compromising the intelligent voice systems. To achieve this goal, hidden voice command~\cite{DBLP:conf/uss/CarliniMVZSSWZ16} utilized inverse MFCC to craft obfuscated commands against ASR systems in a ``black-box'' manner with considerable human effort for obtaining feedbacks. Four different perturbations were introduced to generate noise-like adversarial audios against ASR and SR systems \cite{DBLP:conf/ndss/AbdullahGPTBW19}. Moreover, DophinAttack~\cite{DBLP:conf/ccs/ZhangYJZZX17} was devised to modulate voice commands on inaudible ultrasounds, which can be interpreted by the ASR system, by exploiting the non-linearity of the microphone circuits. However, compared to adversarial example based attacks, these attacks could be easily defended or perceived.

Our initial idea stems from the image-based adversarial attacks. We have overcome particular challenges for generating audio AEs. Compared to the state-of-the-art black-box attacks on acoustic systems~\cite{Taori,Whisper, DBLP:journals/corr/abs-1911-01840,du2019sirenattack}, ours is more generic and practical.

\vspace{-10pt}

% \section{Discussions}
% \bluea{We discuss possible countermeasures to defend against our attacks and potential directions for future research. The detailed discussions can be found in Appendix~\ref{app:discussion}.}

\section{Discussions}\label{app:discussion}
%\subsection{Possible Countermeasures}
We discuss four possible countermeasures to defend against our Occam and NI-Occam below, with detailed performance results of countermeasures in Appendix~\ref{app:Countermeasures} (Tables~\ref{tab:defense123} and~\ref{tab:adversarialtraining}).
% \bluej{We also evaluate the performance of our attacks against various countermeasures, with detailed results in Appendix~\ref{app:Countermeasures} (Tables~\ref{tab:defense123} and~\ref{tab:adversarialtraining}).}

\noindent{\bf Local Smoothing.}
{ Because audio AEs are carefully constructed by adding small perturbations, they can be mitigated by local smoothing.} We can apply a sliding window with the median filter to the adversarial audio signals. Given a data point $x_i$, we replace it with the average of $k$ samples before and after it, \textit{i.e.}, $[x_{i-h}, ..., x_i, ..., x_{i+h}]$. Since the adversarial perturbation is carefully constructed in our attacks, the audio AEs may become less effective after local smoothing transformation. 
\bluej{For example, when $h = 1$, the SRoA of Occam drops from 100\% to 20\%, while the SRoA of NI-Occam drops from 60\% to 40\%.  When $h=3$, all AEs of Occam fail, and NI-Occam remains an SRoA of 40\%. We can find that NI-Occam is more robust to local smoothing. The reason is that the recognized key parts of natural command audios are recovered from audio AEs in NI-Occam, thus making them more robust.}

\noindent{\bf Downsampling.}
{Based on the sampling theory, the high-frequency information in the audios will be lost after the downsampling process, which may disrupt perturbations in the audio AEs.} Thus, audio AEs crafted in our Occam will fail to work after the downsampling and upsampling process, \textit{e.g.}, audio AEs are first downsampled to 12kHz and then upsampled to 16kHz as indicated in our experiments. 
\bluej{Nonetheless, NI-Occam remains a better SRoA,~\ie, 30\% when the downsampling rate (DSR) is 12kHz, indicating that the AEs of NI-Occam have a greater chance of resisting downsampling. }
While downsampling can help to mitigate our attacks, if the dawnsampling/upsampling rates 
are known to the attacker, this countermeasure will be invalid. {This is because the adversary can directly compensate the generated example for the information lost in the downsampling and upsampling process.}

\noindent{\bf Temporal Dependency Based Approach.}
The inherent temporal dependency in audio data was recently utilized in~\cite{DBLP:conf/iclr/YangLCS19} to detect audio AEs due to the disruption of the temporal information. Namely, the first $k$ ($0\sim1$) portions of the whole audio were selected and recognized as $S_k$, and the $k$ portions of the transcription of the whole audio $S_{\{whole, k\}}$ is obtained and compared with $S_k$. By checking the consistency between $S_{\{whole, k\}}$ and $S_k$, audio AEs crafted by Occam can be easily detected. This approach has a strong discriminative ability in identifying AEs targeting at ASR systems, and theoretically it can identify almost all audio AEs against ASR systems.
%More effort is needed to undermine this defense method. But interestingly, when it comes to SR,
Even so, Occam is still effective in attacking SR API services since AEs constructed for SR systems preserve the temporal information like natural audios. For instance, Occam also achieves a success rate of 80\% against Microsoft Azure speaker identification API when our audio AEs were randomly split into two parts, \textit{i.e.}, $k$ and $1-k$ portions of the whole audio.

\bluej{
The defense~\cite{DBLP:conf/iclr/YangLCS19} is better suited for ASR tasks since the temporal dependency is stronger. 
To evaluate the effectiveness of the temporal dependency based approach against (NI-)Occam for ASR, we build a dataset of 40 audio AEs generated using (NI-)Occam with 40 natural command audios, and calculate the detection rates.  For each audio sample, we randomly select $k$ in the range of $[0.2, 0.8]$ and split the audio into two pieces. We then calculate the consistency of the split audios and the whole audio by the word error rate (WER). With a varying WER threshold, we can obtain the ROC curve and finally calculate the area under curve (AUC), where a higher AUC indicates better detection performance. The AUC is 100\% in Occam, showing that the defense can successfully detect all AEs generated by Occam. However, the AUC drops to 68\% when classifying AEs from NI-Occam, which means that NI-Occam is robust to the temporal dependency based approach. This is because the adversarial perturbation generated by NI-Occam usually only occurs in a small piece of audio, and splitting the audio does not disrupt the temporal dependency. Therefore, it is hard for the classifier to detect the AEs generated using NI-Occam by analyzing the temporal information. 
}

\noindent{\bf \bluej{Adversarial Training.}}
\bluej{Adversarial training~\cite{goodfellow2014explaining, DBLP:conf/iclr/MadryMSTV18} is one of the most effective defenses against adversarial attacks in the image domain~\cite{DBLP:conf/ijcai/BaiL0WW21}. The basic idea of adversarial training is to train models on AEs to make the models robust to AEs. It can be formulated as a mini-max optimization problem:
\begin{equation}\label{eq:adv-training}
    \min \limits_{\theta} \mathbb{E}_{(x, y) \sim D} \Big [ \max \limits_{ \Vert\delta \Vert_p <\epsilon}  \mathcal{J}(\theta, x+\delta, y) \Big ],
\end{equation}
}
\bluej{where $\theta$ denotes the deep learning model, $(x, y)$ denotes the original data point, $\delta$ is the adversarial perturbation, and $\Vert\cdot\Vert_p$ is the p-norm. Here, the worst-case samples for the given model are found in the inner maximization problem to train a more robust model via the outer minimization operation.}

\bluej{So far, adversarial training has been extensively studied on image classification tasks \cite{DBLP:conf/iclr/TramerKPGBM18, DBLP:conf/ijcai/CaiLS18, DBLP:journals/corr/abs-1910-08051, DBLP:conf/nips/ShafahiNG0DSDTG19, DBLP:conf/iclr/WongRK20}. However, there is  relatively little research on this defense for speech recognition tasks, which are more complex and challenging. Besides, generating strong audio AEs incurs high computation costs. For example, \cite{CWaudio} reports that it takes approximately one hour to construct an audio AE in a singe NVIDIA 1080Ti GPU, far exceeding the time of generating an adversarial image. 
%Though challenging, it is meaningful to 
%However, unfortunately, since the commercial speech platforms are black-box and closed-source, it is impossible for us to adversarially train the black-box commercial speech models and verify the effect of adversarial training on them.
%Moreover, it is to be noted that the fast-growing training time incurred by adversarial training also presents significant challenges to large commercial models \cite{joshi2021study}, which are also trained on very large voice datasets. 
%Therefore, 
In order to study the effect of adversarial training on ASR systems, we evaluate the performance of adversarial training on the open-source Kaldi. \bluej{For solving the inner maximization problem of Eq. (\ref{eq:adv-training}) that , our targeted NI-Occam is not suitable as it performs the minimization operation, and thus we adopt the untargeted projected gradient
descent (PGD) attack~\cite{DBLP:conf/iclr/MadryMSTV18}, which is usually used in adversarial training, as follows:}
\begin{equation}
    x^{i+1} = x + \mathcal{P}_{\epsilon, p} (x^{i} - x + \alpha \cdot sign(\nabla_{x^{i}} \mathcal{J}(\theta, x^{i}, y) )),
\end{equation}
where $\mathcal{P}_{\epsilon, p}$ is a projection operator on $\mathbf{\emph{L}}_p$ ball, and $\alpha$ the step size.}

\bluej{According to our experimental results on Kaldi (Mini LibriSpeech model\footnote{It is a TDNN based chain Kaldi model trained on Mini-Librispeech dataset. The URL is https://github.com/kaldi-asr/kaldi/tree/master/egs/mini\_librispeech/s5.}) (see Table~\ref{tab:adversarialtraining} in Appendix~\ref{app:Countermeasures}), adversarial training is indeed an effective defense against our targeted NI-Occam attack,~\eg, the SRoA of the AEs drops to 30\% when $\epsilon = 0.002$. But, meanwhile, the WER of the model is increasing from 10.69 to 19.82, which means that the accuracy of the speech recognition drops about 10\%.} \redj{Moreover, with the increase of $\epsilon$, both the SRoA and model accuracy will be significantly reduced. Considering the extreme case when almost all audio AEs fail, the model accuracy drops about 20\%.  And if $\epsilon$ was further increased to eliminate our attack, \eg, approaching $\epsilon = 0.006$, adversarial training is not even able to converge. The reason may be that the audio vector contains many values very close to zero, and they are changed a lot with a high $\epsilon$.}

\bluej{Overall, adversarial training on Kaldi is effective against our attacks. However, while  the SRoA can be reduced by 70\%$\sim$90\%,  it also inevitably brings a significant performance degradation,~\ie, a 10\%$\sim$20\% accuracy loss, and such loss is unacceptable for commercial ASR systems. Moreover, adversarial training will significantly increase the training time and costs~\cite{joshi2021study, DBLP:conf/nips/ShafahiNG0DSDTG19}, particularly for 
\redj{ASR systems. \qi{For example, for the Mini LibriSpeech dataset containing only 5 hours of audio data, it took about 10 days for us to adversarially train Kaldi model on six NVIDIA 2080Ti GPUs, while commonly-used voice datasets, like LibriSpeech and Common Voice, have around 1000 hours of voice data)}, making it almost impractical on large-scale models and datasets.}
\qi{Thus, even with adversarial training our attack cannot be prevented. Therefore, service providers may not be willing to adopt the defense for ``black-box'' commercial models due to the issues above.}}
%besides the large accuracy degradation and high time costs may also hinder service providers to adopt this defense approach on ``black-box'' commercial models.}

\noindent{\bf Remarks.}
We point out several potential research directions to obtain more practical attacks. 
Since the human perception experiment has shown that some audio AEs can be recognized and regarded as abnormal audios, it is necessary to further improve the stealthiness of constructed audio AEs. Moreover, the physical attack against voice control devices in this work will fail when the distance is long or in a very noisy environment. 
\bluej{For example, when we play the AEs at a distance of 50cm from the devices in a noisy cafe, the SRoAs of NI-Occam against Amazon Echo and Apple Siri decrease to 20\%.} 
Thus, it is still challenging to launch physical and black-box attacks on ASR systems at long distances or in noisy environments, which requires more efforts in the future. 
Besides, robust physical adversarial attacks against SR systems in the decision-only and non-interactive settings should also be put on the agenda. Finally, it is also an interesting and meaningful job to achieve black-box adversarial attacks against both ASR and SR systems on one audio AE, because some smart speakers, such as Apple HomePod, will first perform speaker identification on the input audio. 

\section{Conclusion}
\bluea{
In this paper, we proposed two novel black-box adversarial attacks against commercial speech platforms, Occam and NI-Occam. Occam constructs audio AEs against cloud speech APIs in the decision-based black-box setting. 
It is effective under the strictly black-box scenario where the attackers can rely solely on the final decisions. 
Extensive experiments on targeted and untargeted attacks against a wide range of popular open-source and commercial ASR and SR systems demonstrated the effectiveness of Occam. Occam achieves an average SNR of 14.37dB and 100\% SRoA on commercial ASR systems, outperforming the state-of-the-art black-box audio adversarial attacks. 
NI-Occam is the first non-interactive physical attack, simple but effective, which can successfully fool commercial devices without needing any feedback information from the target devices. Extensive experiments showed the effectiveness of the AEs from NI-Occam in attacking Apple Siri, Microsoft Cortana, Google Assistant, iFlytek, and Amazon Echo with an average SRoA of 52\% and SNR of 9.65dB. 
%Finally, we discussed several potential research directions to make our attacks more effective. 
}
%Experiments also showed that, even if only the final decision is exposed to the adversary, existing commercial ASR and SR systems are vulnerable to adversarial attacks, which raises security concerns for developing more robust A(SR) systems. 

%We formulated the decision-based black-box adversarial sample generation as a discontinuous large-scale global optimization problem. We solved it by adaptively decomposing this complicated problem into a set of sub-problems and cooperatively optimizing each one with the CMA-ES by modeling their local geometries. 
%We demonstrated that the cooperative evolution framework is effective in the construction of audio AEs in a divide-and-conquer manner. 

\begin{acks}
This work was partially supported by the National Key R\&D Program of China (2020AAA0107701), NSFC under Grants U20B2049, 61822207, 61822309, 61773310, 62132011, and U1736205, BNRist under Grant BNR2020RC01013, RGC of Hong Kong under Grants CityU 11217819,  CityU 11217620, and R6021-20F, and Laboratory for AI-Powered Financial Technologies. %Qian Wang is the corresponding author.
\end{acks}

\bibliographystyle{ACM-Reference-Format}

\bibliography{references_withoutpages.bib}

% references_withoutpages.bib

\appendix

\section{Adaptive Decomposition Algorithm}\label{sec:AdaptiveDecomposition}
\begin{figure}[t!]
  \centering
  \includegraphics[width=0.95\linewidth]{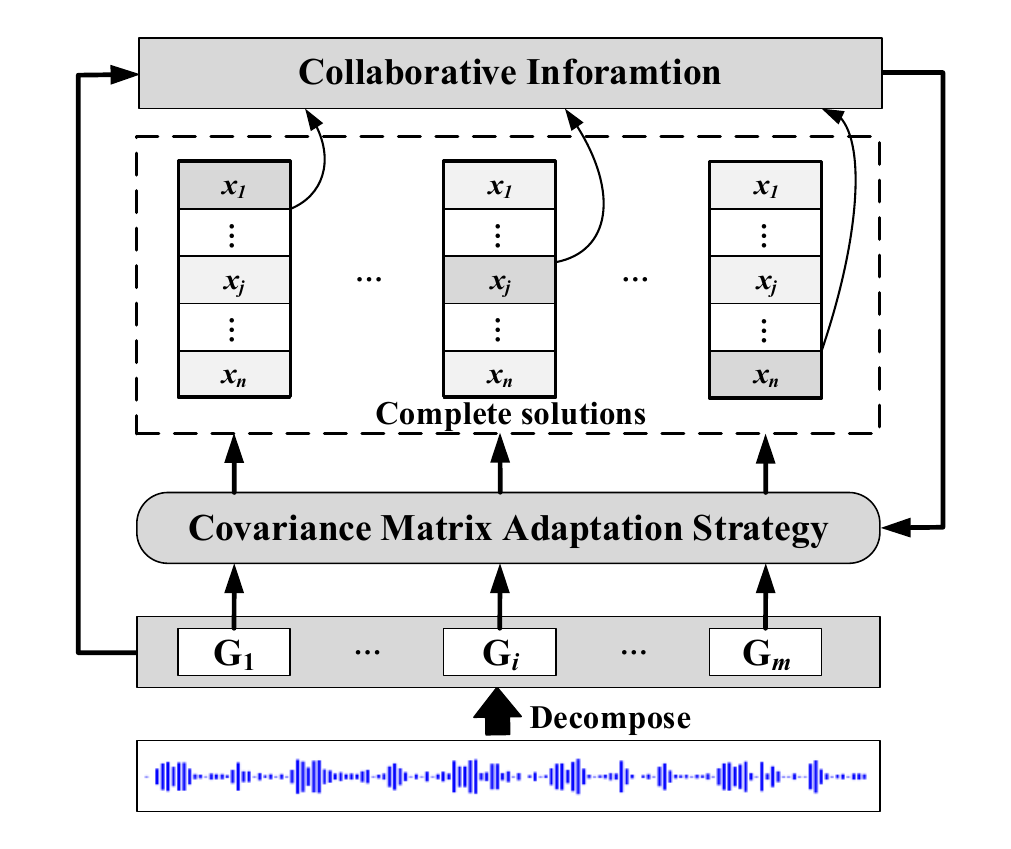} \\
  \caption{Illustration of cooperative co-evolution in Occam.}
  \label{fig:Diagram}
\end{figure}

\bluea{Here we present the detailed adaptive decomposition algorithm in Occam.} {The cooperative co-evolution in Occam is illustrated in Figure~\ref{fig:Diagram}. Intuitively, one may want to probe all possible group sizes and choose the one with the best performance. However, it is very impractical because of the prohibitively high computational overheads. Instead, we propose a dynamic and self-adapting grouping strategy. In our design, we let the group size vary in different stages according to its performance, which is evaluated in a ``pilot'' manner. More specifically, we first divide the variables into subgroups of candidate group sizes and optimize one of the subgroups.
The candidate group size is selected in a gradual manner, \textit{i.e.}, the number of subgroups is twice or half as large as that in the previous stage. We then determine the group size in the next stage according to the performance of pilot test. A candidate size will be adopted in the next stage if the performance of the pilot test is better.
In this way, the grouping strategy greatly reduces computation costs since the pilot test only involves optimizing a small group of variables.
}

The details of the grouping algorithm are described as follows.
Since the degree of interdependence between the variables from the natural audio and the audio AE gradually increases, we first initialize the number of groups as $m=1$.

\begin{enumerate}[1)]
\item Create two lists $R = \{r_1, r_2, r_3, r_4\}$  and $\Delta = \{\delta_{m/2}, \delta_{m}, \delta_{2m}\}$ to record the impact of the four grouping strategies in the candidate pool and different group sizes, respectively. Initialize  $R = 0$, $\Delta = 0$ and $m = 1$.
\item Compute the probabilities of selecting each grouping strategy in the candidate pool as
\begin{equation}
      prob_{j} = \frac{e^{r_j}}{\sum_{j = 1}^4{e^{r_j}} }, j = {1,2,3,4}.
\end{equation}
\item Select a grouping strategy according to probabilities $\{prob_j\}_{j=1}^4$.
\item Group the variables into $m$ subgroups $\mathcal{G}_{m}=\{G_1, \dots, G_{m}\}$ according to the selected grouping strategy.
\item Optimize the subspaces on $\mathcal{G}_{m}$ until the current stage ends. Update $r_i$ and $\delta_{m}$ as
\begin{equation}
r_j = \delta_{m} = \left |\frac{(v-v')}{v\cdot m}\right|,
\end{equation}
where $v$ is the best fitness of the last cycle, and $v'$ is the best fitness of the current cycle.
\item Run pilot test.
    \begin{enumerate}[i.]
    \item If $\delta_{m/2} = 0$ and $m/2 \geq 1$, group the variables into $m/2$ subgroups $\mathcal{G}_{m/2}=\{G_1, \ldots, G_{m/2}\}$ according to the selected grouping strategy, and optimize $G_1$ to obtain the performance record
    \begin{equation}
     \delta_{m/2}  = \left |\frac{(v-v')}{v}\right|.
    \end{equation}
    \item If $\delta_{2m} = 0$ and $2m \leq n$, group the variables into $2m$ subgroups $\mathcal{G}_{2m}=\{G_1, \ldots, G_{2m}\}$ according to the selected grouping strategy, and optimize $G_1$ to obtain the performance record
    \begin{equation}
     \delta_{2m} = \left |\frac{(v-v')}{v}\right|.
    \end{equation}
    \end{enumerate}
\item If $\delta_{m/2}$ is the maximum among $\Delta$, then set $m = m/2$, $\delta_{2m} = \delta_{m}, \delta_{m} = \delta_{m/2}$, and $\delta_{m/2} = 0$. Else if $\delta_{2m}$ is the maximum, then set $m = 2m$, $\delta_{m/2} = \delta_{m}, \delta_{m} = \delta_{2m}$, and $ \delta_{2m} = 0$.
\item  Go to step 3.
\end{enumerate}

%\reda{[1. max among who? Is $\delta_{m}$ involved in the comparison? 2. $m/2$ or $2m$? Why ``or''? Is this mean that we should try both $m/2$ and $2m$? If not, how to choose $m/2$ or $2m$?]}

\section{Differential Evolution}\label{app:DEA}
Differential evolution is a gradient-free optimization method like CMA-ES. To sufficiently justify the design choice of CMA-ES, we also include the evaluations on the differential evolution attack (DEA). Our experimental results (see Tables~\ref{tab:deepspeech} and~\ref{tab:apisnr}) show that DEA cannot perform well on audio data. 
Here we present the algorithm of differential evolution, as shown in Alg.~\ref{alg:DEA}. To more easily generate offspring that is closer to the original audio, a bias term is also added into the mutation step, and the crossover step is removed.

\begin{algorithm}[!t]
    \caption{Differential Evolution Attack}\label{alg:DEA}
    \begin{algorithmic}[1]
        \REQUIRE {The original audio $x$, the initial adversarial audio $x'$, the input space dimension $n$, the Population $NP$, and the total number of iterations $G$.}
        \ENSURE The adversarial audio sample $x^*$.
        \STATE Initialize the following parameters:
        \begin{itemize}
        \item $NP=10$, $G=3000$, $F=0.5$, 
        \item $g=0$, $\sigma = 0.003, \mu = 0.2$.
        \end{itemize}
        \FOR {$i$ = $0$ to $NP$}
            \STATE Sample $z$ $\sim$ $\mathcal{N}(0, \sigma^{2})$;
            \STATE $X^0_i=x^{'}+z$;
        \ENDFOR
        \WHILE{$g<G$}
            \STATE $\triangleright$ Mutation
            \FOR {$i$ = $0$ to $NP$}
                \STATE Generate three random integers $r_1$, $r_2$ and $r_3$ $\in$ $\left\{ 0,1,...,NP-1  \right\}$, with {$r_1$ $\neq$ $r_2$ $\neq$ $r_3$};
                \STATE $Xm_i=X^g_{r_1}+ rand(0,\mu)(x-X^g_{r_1})+F \times (X^g_{r_2}-X^g_{r_3})$;
            \ENDFOR
            \STATE $\triangleright$ Selection
            \FOR {$i$ = $0$ to $NP$}
                \IF {$\mathcal{L}(Xm_i)$ $\textless$ $\mathcal{L}(X^g_i)$}
                    \STATE $X^{g+1}_i=Xm_i$;
                \ENDIF
            \ENDFOR
            \STATE $g \gets g + 1$;
        \ENDWHILE
        \RETURN {$x^*$}.
    \end{algorithmic}
\end{algorithm}

\section{Detailed Experiment Setting}\label{app:detailedsetting}

\subsection{Datasets}\label{app:dataset}
\textbf{Common Voice}. Common Voice~\cite{commonvoice} is the largest open-source human voice dataset launched by Mozilla, including 28 different languages and nearly 1,800 validated hours of voice data. The Common Voice dataset is used to train ASR systems and test the effectiveness of audio AEs.

\noindent\textbf{Song}. Song is released by CommanderSong~\cite{CommanderSong} and also used in Devil's Whisper \cite{Whisper}. It contains 20 songs divided into four categories on average, including the soft, popular, rock, and rap. Similar to Devil's Whisper \cite{Whisper}, we select a total of 10 songs in the soft and popular categories, which are less noisy and easier to disguise, to evaluate the performance of Occam against speech recognition APIs.

\noindent\textbf{LibriSpeech}.
LibriSpeech~\cite{panayotov2015librispeech} is an English speech corpus that consists of 1,000 hour reading audio sampled at 16kHz.

\noindent\textbf{VoxCeleb}. VoxCeleb~\cite{DBLP:conf/interspeech/NagraniCZ17} is an open-source large-scale audio dataset that contains more than 100,000 utterances from more than 7,000 celebrities. It consists of short human voice clips extracted from interview videos uploaded to YouTube. In our experiments on the SR systems, we use part of VoxCeleb as the enrollment data.

\subsection{Target Systems}\label{app:targetsystem}
\textbf{DeepSpeech.} DeepSpeech~\cite{deepspeech} is a state-of-the-art ASR system that performs speech-to-text tasks. It is usually used as the target model in adversarial attacks~\cite{KAM2018,Taori}. Note that although DeepSpeech is open-source, Occam does not require any internal knowledge about the target model.

\noindent\textbf{Commercial Cloud Speech APIs.}
Speech-to-text API services are widely used in many applications. To evaluate Occam on commercial services, we choose six popular API services, including Google Cloud Speech-to-Text~\cite{Google}, Microsoft Azure Speech Service~\cite{Microsoft}, Alibaba Short Speech Recognition~\cite{Alibaba}, Tencent Short Speech Recognition~\cite{Tencent}, and iFlytek voice dictation~\cite{iFlytek}. For the Google's Speech-to-Text API, we select the ``command\_and\_search model'' as the target models. The characteristics of these APIs are listed in Table~\ref{tab:api}.
Note that although some APIs provide confidence scores, our attack does not require such information.

\noindent\textbf{Commercial Speaker Recognition Systems.}
We test Occam on the Microsoft Azure's speaker recognition system and the Jingdong speaker recognition system~\cite{Jingdong}. Microsoft Azure's API~\cite{Microsoft} can perform speaker identification (who is the speaker) and speaker verification (whether the speaker is legal). It only returns the decision (\ie, the predicted speaker) along with three confidence levels (\ie, low, normal, or high). Jingdong's API can perform speaker verification and only returns the final result,~\ie, accept or reject.

\noindent\textbf{Commercial Voice Control Devices.}
\bluea{
We test NI-Occam on five commercial voice control devices,~\ie, Apple Siri, iFlytek, Microsoft Cortana, Google Assistant, Amazon Echo. In our experiments, Apple Siri, iFlytek, Microsoft Cortana, Google Assistant are applications installed in on-the-shelf smartphones, and Amazon Echo is an intelligent voice-controlled speaker.  
}

\subsection{Hardware}
In our experiments, we conduct the attacks against DeepSpeech on a server equipped with four Nvidia 2080Ti GPUs, a six-core Intel(R) Xeon(R) W-2133 CPU 3.60GHz, 62 Gigabyte RAM, and a Hard Drive with 1.37 Terabytes. For the experiment on speech API services, we adopt several laptops, including a Lenovo ThinkPad X1 Carbon 4th with Core Intel i5-6200U CPU 2.30GHz and 8 Gigabyte RAM, a Microsoft Corporation Surface Pro 6 with Core Intel i7-8650U CPU 1.90GHz and 8 Gigabyte RAM, and a Lenovo ThinkPad X1 Carbon 5th with Core Intel i7-7500U CPU 2.70GHz and 8 Gigabytes RAM. Besides, we attack the voice assistant including Apple Siri with version 13.6.1 and iFlytek with version 10.0.8 on an iPhone 11, Cortana with version 3.3.3 on a Samsung C9000 with version 3.3.3, and Google Assistant with version 2.5.1 on a Nokia 7 plus. We play the audio AEs using a JBL Clip 3 portable speaker.

%(Google Assistant) on a phone (Samsung SM-G9730) by using the adversarial audio played out by an iPad Air 2.

\begin{table*}[t]
\centering
\caption{\bluea{Experimental results on targeted attacks against DeepSpeech.}}
\resizebox{\textwidth}{15mm}
{
\begin{tabular}{c|cc|cc|cc|cc|cc|cc|cc|cc} 
\hlinew{1.2pt}
\multirow{2}{*}{\textbf{\textbf{Dataset}}}  & \multicolumn{2}{c|}{\textbf{GAA}} & \multicolumn{2}{c|}{\textbf{SGEA}} & \multicolumn{2}{c|}{\begin{tabular}[c]{@{}c@{}}\textbf{Boundary} \\\textbf{Attack}\end{tabular}} & \multicolumn{2}{c|}{\textbf{Opt-Attack}} & \multicolumn{2}{c|}{\begin{tabular}[c]{@{}c@{}}\textbf{Evolutionary} \\\textbf{Attack}\end{tabular}} & \multicolumn{2}{c|}{\textbf{HSJA}} & \multicolumn{2}{c|}{\textbf{DEA}} & \multicolumn{2}{c}{\textbf{Occam}}  \\ 

\cline{2-17}
     & SRoA & SNR               & SRoA & SNR                & SRoA  & SNR                          & SRoA  & SNR                     & SRoA  & SNR                              & SRoA  & SNR               & SRoA  & SNR                                 & SRoA  & SNR                     \\ 
\hlinew{1.2pt}
\textbf{Songs}          & 1/10 & 23.78             & 4/10 & 21.62              & 10/10 & 8.03                         & 10/10 & 2.88                    & 10/10 & 1.20                              & 10/10 & 5.88              & 10/10 & -0.26                               & \textbf{10/10} & \textbf{12.86}                   \\ 
\hline
\textbf{LibriSpeech}    & 1/10 & 14.72             & 7/10 & 14.60              & 10/10 & 4.80                         & 10/10 & 2.93                    & 10/10 & 1.09                             & 10/10 & 1.94              & 10/10 & -1.41   & \textbf{10/10} & \textbf{11.30}                   \\ 
\hline
\textbf{Common Voice}   & 1/10 & 17.69            & 8/10 & 19.14              & 10/10 & 3.41                         & 10/10 & 2.01                    & 10/10 & 0.25                             & 10/10 & 1.78              & 10/10 & -0.52                               & \textbf{10/10} & \textbf{13.80}                   \\
\hlinew{1.2pt}
\textbf{Average}      & 1/10                                       & 18.73                                       & 6.3/10               & 18.45               & 10/10                                         & 5.41                                          & 10/10            & 2.61            & 10/10           & 0.85            & 10/10                                        & 3.2                                       & 10/10       & -0.73                    & \textbf{10/10}                                       & \textbf{12.65} \\
\hlinew{1.2pt}
\end{tabular}
}\label{tab:deepspeech}
\end{table*}

\section{Supplementary Evaluation of Occam}
\subsection{Evaluation on Open-source ASR Systems}\label{app:deepspeech}
We evaluate the effectiveness of the audio AEs generated by the attacks in Table~\ref{tab:deepspeech}. 
These AEs are generated after 200,000 queries (100,000 times on Song\footnote{The original data from Song requires fewer queries because the signal power of the songs is relatively strong.}) on DeepSpeech. 
%GAA and SGEA require the knowledge of logits (prediction score) inside the model. 
Since Genetic Algorithm-based Attack (GAA)~\cite{Taori} and Selective Gradient Estimation Attack (SGEA)~\cite{TIFS-SGEA} need to leverage the prediction scores of the model, the AEs achieve higher SNRs than other decision-based attacks. However, neither of them can successfully attack DeepSpeech with a 100\% SRoA-ASR. 
In the other six decision-based attacks, the initial example is originally adversarial and then trained to approach the target example. Therefore, the SRoA of these attacks is always 100\%. However, it is worth noting that SRoA is not the only metric that determines whether an AE can successfully attack the target system. An effective AE needs to fool both the model and the human, where SNR must be used as another important factor to measure the effectiveness of AEs. Although the six decision-based attacks can all achieve a 100\% SRoA-ASR, only the AEs generated by our attack obtain high SNRs (with the best SNR of 13.80dB).

\begin{figure}[!t]
    \centering
    \subfigure[Common Voice]{
    \begin{minipage}[t]{0.48\linewidth}
        \centering
        \includegraphics[width=\linewidth]{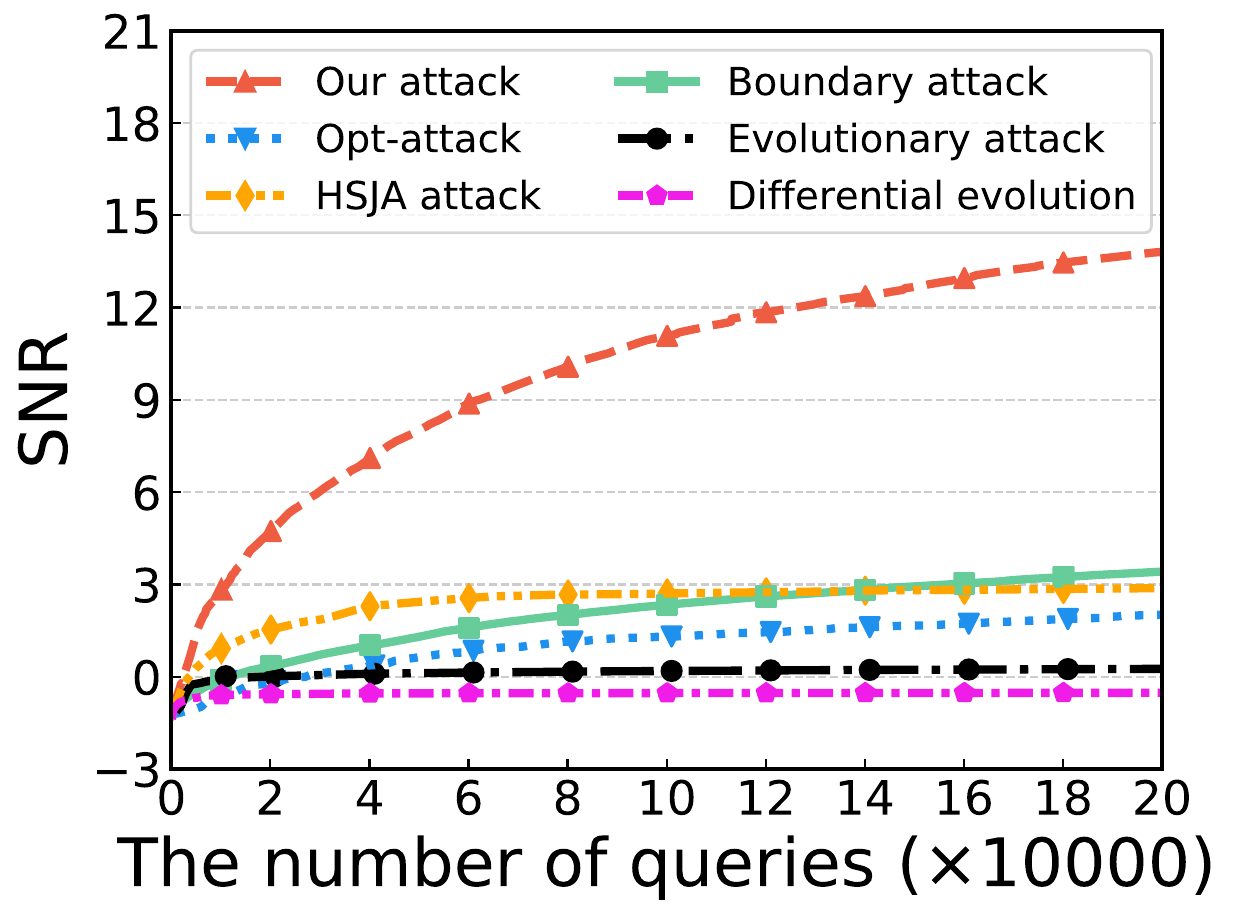}
        \vspace{-3mm}
        %\caption{Continuous optimization}
        \label{fig:commonvoicesnr}
    \end{minipage}}
    \subfigure[Song]{
    \begin{minipage}[t]{0.48\linewidth}
        \centering
        \includegraphics[width=\linewidth]{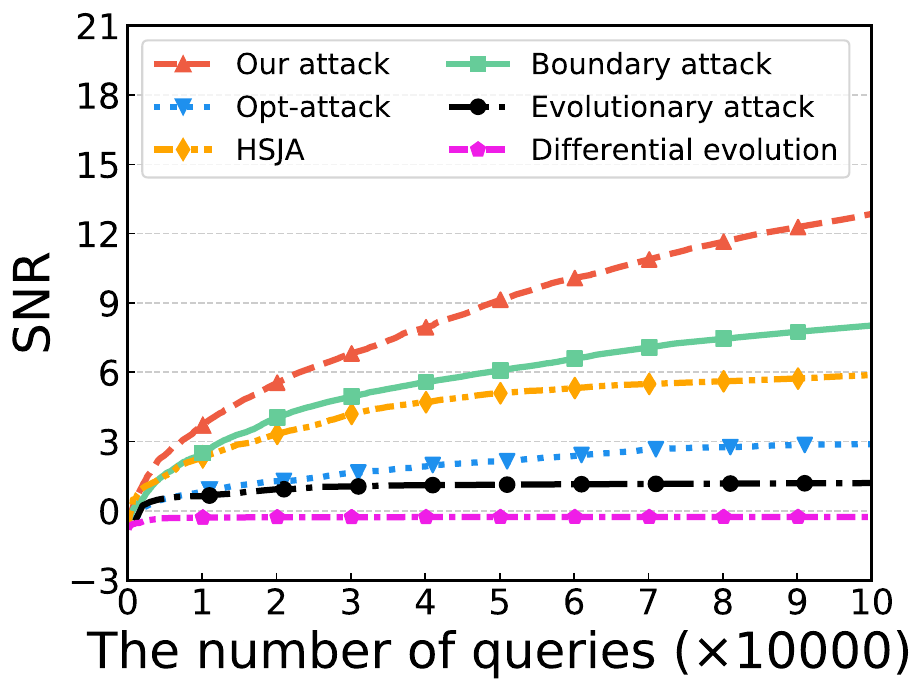}
        \vspace{-3mm}
        %\caption{Discontinuous optimization}
        \label{fig:songsnr}
    \end{minipage}}
    \vspace{-2mm}
    \caption{\bluea{SNRs of the audio AEs after querying DeepSpeech for different times over (a) Common Voice and (b) Song.}}
    \label{fig:snrquery}
\end{figure}

To evaluate the efficiency of Occam, we tested the SNRs of the AEs generated after different numbers of queries on DeepSpeech.
As shown in Figure~\ref{fig:snrquery}, Occam achieves a high SNR within a small number of queries. Note that the dataset Song requires fewer queries because the audio in Song is less noisy and more powerful, making it easier to generate an audio AE with a high SNR.
\bluea{We also find that the growth rate of SNR significantly decreases with the increase of SNR. 
For example, the SNR of our AEs can quickly converge to 12.86dB, while the evolutionary attack may require hundreds of thousands of queries to increase SNR from 1.20dB to 12.86dB.
Hence, the results show that the co-evolution algorithm has a higher bound of SNR and a faster convergence rate than the evolution algorithm on audio data, validating the effectiveness of the CC framework in constructing audio AEs.}

 \begin{figure}[t!]
  \centering
  \includegraphics[width=0.95\linewidth]{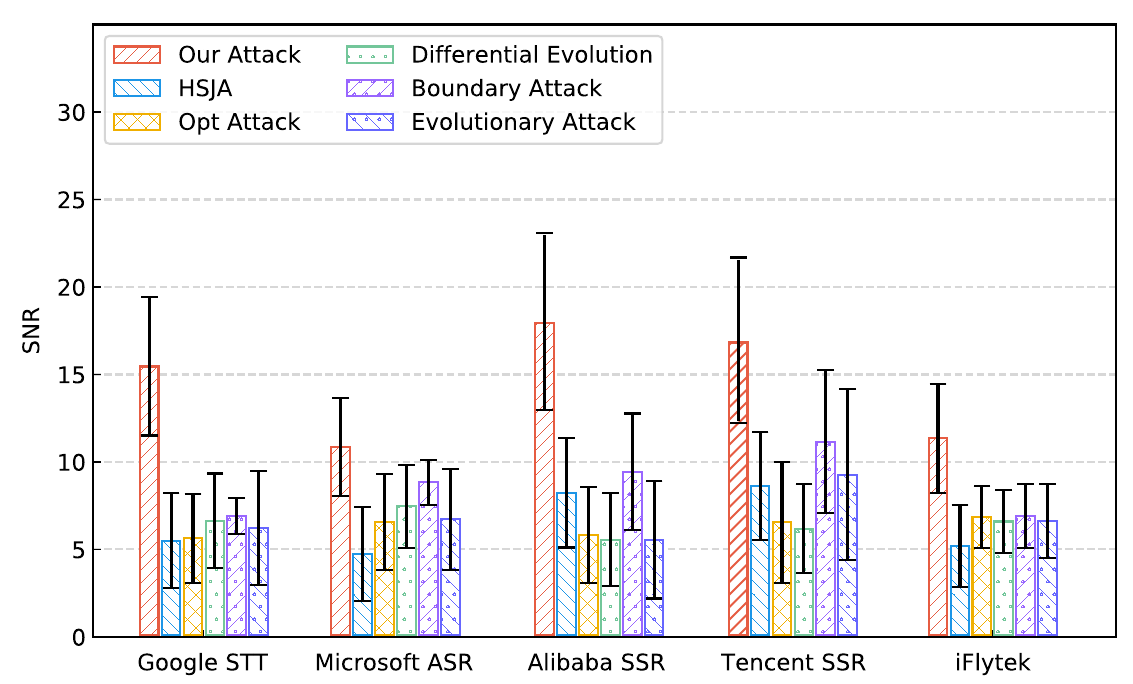} \\
  \vspace{-2mm}
  \caption{\bluea{SNRs of the AEs from targeted attacks against the commercial services.}}
  \label{fig:confidence}
\end{figure}

\begin{table*}[!t]\small
% \centering
\caption{Experimental results of untargeted attacks on commercial cloud speech-to-text APIs.}
\begin{tabular}{c|c|c|c|c|c|c}
\hlinew{1.2pt}
\textbf{ Model  }        & \textbf{ Boundary Attack  } & \textbf{ Opt-Attack } & \textbf{ Evolutionary Attack } & \textbf{HSJA} & \textbf{DEA} & \textbf{ Occam }  \\
\hlinew{1.2pt}
\textbf{Google STT}        & 26.43                       & 19.63                 & 31.20                 &30.04 & 16.31      & \textbf{31.43}                 \\
\hline
\textbf{Microsoft ASR } & 15.43                       & 11.89                 & 27.37                     & 19.60 & 11.08  & \textbf{28.68}                  \\
\hline
%\textbf{IBM STT}       & 25.15                      & 28.54                 & \underline{\textbf{41.30}}             &  32.62& 26.32            & \textbf{38.97}                  \\\hline
\textbf{Alibaba SSR}   & 17.84                       & 15.05                 & 27.85                        &20.66&17.49  & \textbf{29.57}                 \\
\hline
\textbf{Tencent SSR}   & 20.38                       & 17.02                 &30.16                    &21.51&18.03  &\textbf{31.68}                  \\
\hline
\textbf{iFlytek}       &42.66                       & 29.07                 & 38.59             &39.82&26.61             & \textbf{40.46}                \\
\hlinew{1.2pt}
\end{tabular}\label{tab:untarget}

\begin{tablenotes}
\item    \footnotesize{
Note that, (i) this table shows the SNRs of the audio AEs generated after 200 queries.
(ii) This table does not show the success rate because all of the six attacks achieve a success rate of 100\%.
(iii) The best and the second best results on each API are marked in bold with underling and bold font, respectively.
 }
  \end{tablenotes}

% \footnotesize{Note that, (1) This table shows the SNRs of the audio AEs generated after 200 queries.

% (2) This table does not show the success rate because all of the six attacks achieve a success rate of 100\%.

% (3) The best and the second best results on each API are marked in bold with underling and bold font, respectively.
% }

\end{table*}

\subsection{Evaluation on Cloud Speech APIs}\label{app:wave}
\bluea{
For completeness, we give some additional experimental results as a complement to Section~\ref{sec:api}.
}

\begin{figure}[t!]
  \centering
  \includegraphics[width=\linewidth]{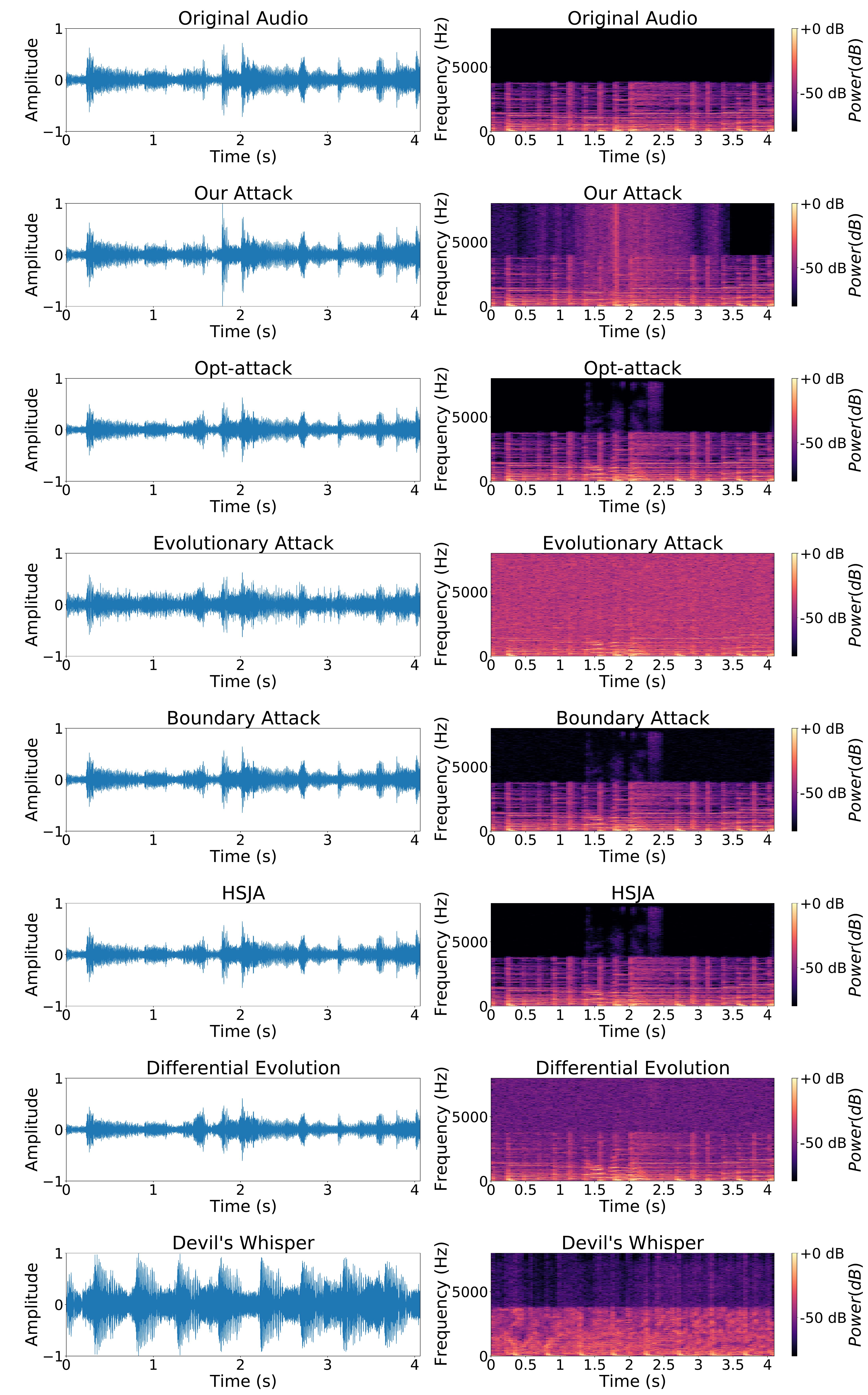} \\
  \vspace{-2mm}
  \caption{\bluea{Waveforms and spectrograms of the original audio and adversarial audios generated by the targeted attacks against Alibaba SSR. The audio AEs can be recognized as ``call my wife''. }}
  \label{fig:specfull}
\end{figure}

We present Figure~\ref{fig:confidence} to show the SNRs with confidence intervals of 68.2\% (-std, +std). The results show that Occam has the highest SNRs among all the decision-based attacks.
Figure~\ref{fig:specfull} shows the waveforms and spectrograms of the original audio and the adversarial audios generated by the seven attacks.
It can be seen that the waveforms of the original audio and the audio AEs generated by Occam are almost the same. However, the differences in other attacks are more noticeable and thus more likely to be perceived by humans.
%The results of the attacks against ASR and SR systems with different queries are shown in Table~\ref{tab:apisnrfull} and Table~\ref{tab:srfull}, respectively. Occam outperforms other attacks even with a smaller number of queries.

As for the untargeted attacks,
the results in Table~\ref{tab:untarget} show that the SNRs of AEs generated from untargeted attacks are much higher than those from targeted attacks. 
In this experiment, the performance of the evolutionary attack is comparable to Occam. Since the implementation of untargeted attacks is much simpler than the targeted attack (\eg, 200 queries are enough), the optimization problem on untargeted attacks can be easily solved without being decomposed into a set of simpler sub-problems. Thus, approaches without a cooperative co-evolution framework can also work well. However, concerning the large-scale and complex problems of generating targeted audio AEs, these approaches become ineffective. 
 \bluea{Besides, Devil's Whisper does not explicitly support untargeted attacks. To launch the untargeted attack by using the methodology of Devil's Whisper, the attacker needs to intentionally set a wrong phrase as the target phrase. This step incurs a large amount of queries.} In contrast, our methodology can successfully generate effective audio AEs within 200 queries.

\subsection{Evaluation on Human Perception}\label{app:human}
Similar to the human perception experiments on NI-Occam, we also evaluate the performance of Occam about the human perception. The experiment settings are the same as those in Section~\ref{sec:human}.

For each commercial service, the volunteers need to listen to 10 pieces of audio AEs generated from the boundary attack, the opt-attack, the evolutionary attack, HSJA, DEA, Devil's Whisper, and Occam. The audio AEs are generated by querying the target systems for 200,000 times with original audios from the dataset Song. In all, we get 37$\times$10 samples from each attack on each commercial service (a total of 37$\times$10$\times$6 samples for one attack). Note that, since Devil's Whisper does not achieve a 100\% SRoA, the volunteers only rate the successful AEs. Each volunteer ranks 3, 5, 7, 8, and 4 successful AEs from Devil's Whisper on Alibaba, Google, iFlytek, Microsoft, and Tencent, respectively. 

Table~\ref{tab:human} presents the results of the experiments on human perception. 
Overall, Occam has the highest rate of ``normal''. 
More than 70\% volunteers think the audio generated from Occam is ``normal'' or ``noise'', which is comparable to Devil's Whisper. However, about 50\% AEs from Devil's Whisper fail to fool the ASR systems in the first place. The results show that Occam is more effective in fooling both the model and the human than other possible attacks.

\begin{table*}[p]
	\centering
	\caption{Evaluation results on human perception of different attacks in the digital world.}
	\renewcommand\arraystretch{1.15}
	\begin{tabular}{c|c|c|c|c|c|c} 
		\hlinew{1.2pt}
		\textbf{Commercial Service}             & \textbf{Method}     & \textbf{Normal (\%)} & \textbf{Noise (\%)} & \textbf{Talking (\%)} & \begin{tabular}[c]{@{}c@{}}\textbf{Once-}\\\textbf{recognize (\%)}\end{tabular} & \begin{tabular}[c]{@{}c@{}}\textbf{Twice-}\\\textbf{recognize (\%)}\end{tabular}  \\ 
		\hlinew{1.2pt}
		\multirow{7}{*}{\textbf{Alibaba SSR}}   & Boundary Attack     & 26.76                & 23.24               & 50.00                 & 21.35                                                                         & 28.92                                                                             \\ 
		\cline{2-7}
		& DEA                 & 6.76                 & 23.24               & 70.00                 & 45.14                                                                           & 52.70                                                                             \\ 
		\cline{2-7}
		& Evolutionary Attack & 2.70                 & 32.16               & 65.14                 & 43.24                                                                           & 46.76                                                                             \\ 
		\cline{2-7}
		& HSJA                & 8.65                 & 37.30               & 54.05                 & 31.62                                                                           & 36.22                                                                             \\ 
		\cline{2-7}
		& Opt-Attack          & 4.59                 & 29.73               & 65.68                 & 38.65                                                                           & 43.24                                                                             \\ 
		\cline{2-7}
		& Devil's Whisper     & 8.11                & 51.35               & 40.54                 & 14.41                                                                           & 18.02                                                                             \\ 
		\cline{2-7}
		& Occam          & 50.54                & 35.41               & 14.05                 & 3.24                                                                            & 6.22                                                                              \\ 
		\hlinew{1.2pt}
		\multirow{7}{*}{\textbf{Google STT}}    & Boundary Attack     & 4.59                 & 21.89               & 73.51                 & 35.68                                                                           & 39.73                                                                             \\ 
		\cline{2-7}
		& DEA                 & 0.54                 & 21.35               & 78.11                 & 34.59                                                                           & 42.70                                                                             \\ 
		\cline{2-7}
		& Evolutionary Attack & 3.51                 & 17.30               & 79.19                 & 33.24                                                                           & 42.16                                                                             \\ 
		\cline{2-7}
		& HSJA                & 5.14                 & 15.68               & 79.19                 & 37.03                                                                           & 45.68                                                                             \\ 
		\cline{2-7}
		& Opt-Attack          & 6.76                 & 12.70               & 80.54                 & 40.00                                                                           & 46.22                                                                             \\ 
		\cline{2-7}
		& Devil's Whisper     & 18.38                & 60.54              & 21.08                 & 7.03                                                                            & 8.11                                                                              \\ 
		\cline{2-7}
		& Occam          & 40.00                & 38.38               & 21.62                & 6.22                                                                            & 14.59                                                                             \\ 
		\hlinew{1.2pt}
		\multirow{7}{*}{\textbf{iFlytek}}       & Boundary Attack     & 3.51                 & 23.78               & 72.70                 & 33.24                                                                           & 43.51                                                                            \\ 
		\cline{2-7}
		& DEA                 & 0.00                 & 23.78               & 76.22                 & 34.59                                                                           & 41.08                                                                             \\ 
		\cline{2-7}
		& Evolutionary Attack & 5.14                 & 23.24               & 71.62                 & 33.24                                                                           & 39.73                                                                             \\ 
		\cline{2-7}
		& HSJA                & 1.62                 & 20.27               & 78.11                 & 35.68                                                                           & 40.70                                                                             \\ 
		\cline{2-7}
		& Opt-Attack          & 3.51                 & 19.73               & 76.76                 & 37.03                                                                           & 43.24                                                                             \\ 
		\cline{2-7}
		& Devil's Whisper     & 17.76                & 48.26               & 33.98                 & 8.11                                                                            & 13.13                                                                             \\ 
		\cline{2-7}
		& Occam          & 22.16                & 41.08               & 36.76                 & 14.05                                                                           & 21.08                                                                            \\ 
		\hlinew{1.2pt}
		\multirow{7}{*}{\textbf{Microsoft ASR}} & Boundary Attack     & 13.51                & 25.95               & 60.54                 & 27.57                                                                           & 29.19                                                                             \\ 
		\cline{2-7}
		& DEA                 & 8.11                 & 23.24               & 68.65                 & 31.08                                                                           & 37.57                                                                             \\ 
		\cline{2-7}
		& Evolutionary Attack & 1.08                 & 23.78               & 75.14                 & 36.22                                                                           & 44.59                                                                             \\ 
		\cline{2-7}
		& HSJA                & 1.08                 & 17.84               & 81.08                 & 39.19                                                                           & 47.03                                                                             \\ 
		\cline{2-7}
		& Opt-Attack          & 4.05                 & 19.73              & 76.22                 & 36.76                                                                           & 43.51                                                                             \\ 
		\cline{2-7}
		& Devil's Whisper     & 29.05                & 45.27               & 25.68                 & 8.78                                                                            & 13.85                                                                             \\ 
		\cline{2-7}
		& Occam          & 31.08                & 28.92               & 40.00                 & 16.22                                                                           & 24.05                                                                             \\ 
		\hlinew{1.2pt}
		\multirow{7}{*}{\textbf{Tencent SSR}}   & Boundary Attack     & 29.73                & 28.11               & 42.16                 & 21.62                                                                           & 29.19                                                                             \\ 
		\cline{2-7}
		& DEA                 & 1.08                 & 20.27               & 78.65                 & 34.59                                                                           & 37.57                                                                             \\ 
		\cline{2-7}
		& Evolutionary Attack & 14.59                & 33.24               & 52.16                 & 28.65                                                                           & 31.08                                                                             \\ 
		\cline{2-7}
		& HSJA                & 18.65                & 30.81               & 50.54                 & 29.19                                                                           & 32.16                                                                             \\ 
		\cline{2-7}
		& Opt-Attack          & 11.62                 & 19.73               & 68.65                 & 35.68                                                                           & 41.62                                                                             \\ 
		\cline{2-7}
		& Devil's Whisper     & 12.84                & 37.84               & 49.32                 & 10.14                                                                           & 15.54                                                                             \\ 
		\cline{2-7}
		& Occam          & 45.41                & 34.05               & 20.54                 & 5.41                                                                            & 7.84                                                                              \\ 
		\hlinew{1.2pt}
		
		\multirow{7}{*}{\textbf{Average}}       & Boundary Attack     & 15.62                & 24.59               & 59.78                 & 27.89                                                                           & 34.11                                                                             \\ 
		\cline{2-7}
		& DEA                 & 3.30                 & 22.38               & 74.32                 & 36.00                                                                           & 42.32                                                                             \\ 
		\cline{2-7}
		& Evolutionary Attack & 5.41                 & 25.95               & 68.65                 & 34.92                                                                           & 40.86                                                                             \\ 
		\cline{2-7}
		& HSJA                & 7.03                 & 24.38               & 68.59                 & 34.54                                                                          & 40.76                                                                             \\ 
		\cline{2-7}
		& Opt-Attack          & 6.11                 & 20.32               & 73.57                 & 37.62                                                                          & 43.57                                                                            \\ 
		\cline{2-7}
		& Devil's Whisper     & 17.23                & 48.65               & 34.12                & 9.69                                                                           & 13.73                                                                             \\ 
		\cline{2-7}
		& Occam          & 37.84                & 35.57               & 26.59                 & 9.03                                                                           & 14.76                                                                             \\
		\hlinew{1.2pt}
	\end{tabular}\label{tab:human}
	
	\
	
	\begin{tablenotes}
\item    \footnotesize{
Note that, 
(i) ``Normal'' means that the volunteer regards the audio as a normal audio. 
(ii) ``Noise'' means that the volunteer can feel some noises. 
(iii) ``Talking'' means that the volunteer can hear talking in the audio. If the volunteer thinks there is talking in the audio, he/she is then asked to recognize the content of the talking. 
(iv) The audio will be labeled as ``once-recognize'' or ``twice-recognize'' if the volunteer recognizes over half of the content after listening to the audio once or twice, respectively.
 }
  \end{tablenotes}
\end{table*}

%\noindent\textbf
\section{A Human Study on Audio AEs with Different SNRs}\label{app:SNR-human}
\bluej{Recall that we use two important metrics to evaluate the effectiveness of the audio AEs,~\ie, the SRoA (Success Rate of Attack) and SNR. 
Both metrics are crucial to determine whether an AE can successfully attack the target system, since an effective AE needs to fool both the model and the human simultaneously.
To demonstrate the necessity of SNR as an evaluation criterion, we conduct a small human study to help understand how the metric of SNR affects the effectiveness and quality of an audio AE.}

\bluej{
In this experiment, we surveyed 30 volunteers who are sensitive to sound, including 17 males and 13 females. Similar to the human perception experiment (See Section~\ref{sec:human}), we first show some examples of ``noise'', ``normal'', ``talking'', and ``recognized'', and then ask the volunteers to listen to the audio AEs and tell their views about them. 
The audio AEs are generated from Occam against Alibaba and Tencent, with SNRs ranging from 6dB to 16dB. Specifically, we generate 10 AEs for each SNR group.
As shown in Table~\ref{tab:snrhuman}, when the SNR increases (\ie, the noise becomes smaller), more volunteers consider the AE as a normal one. Specifically, when the SNR is lower than 8dB, over 84\% of volunteers can hear talking in the AE audios, and more than 31\% of volunteers can recognize the commands in the AEs. The results validate that if the SNR is low, it is easy for the users to notice or even recognize the AEs. Since the audio AEs need to be stealthy enough to fool the human, AEs with high SNRs are considered more in practice.   
}

\begin{table}[!t]\small
\centering
\caption{\bluej{Evaluation results on human perception of AEs with different SNRs.}}
\small
\begin{tabular}{c|c|c|c|c|c} 
\hlinew{1.2pt}
\begin{tabular}[c]{@{}c@{}}\textbf{SNR}\\\textbf{(dB)}\end{tabular} & \begin{tabular}[c]{@{}c@{}}\textbf{Normal}\\\textbf{(\%)}\end{tabular} & \begin{tabular}[c]{@{}c@{}}\textbf{Noise} \\\textbf{(\%)}\end{tabular} & \begin{tabular}[c]{@{}c@{}}\textbf{Talking} \\\textbf{(\%)}\end{tabular} & \begin{tabular}[c]{@{}c@{}}\textbf{Once-}\\\textbf{recognize} \textbf{(\%)}\end{tabular} & \begin{tabular}[c]{@{}c@{}}\textbf{Twice-}\\\textbf{recognize} \textbf{(\%)}\end{tabular}  \\ 
\hlinew{1.2pt}
\textbf{16}                                                & 36.00                                                & 38.33                                               & 25.67                                                  & 3.33                                                         & 4.33                                                            \\ 
\hline
\textbf{14}                                                & 27.67                                                & 35.33                                               & 37.00                                                  & 6.67                                                         & 7.33                                                            \\ 
\hline
\textbf{12}                                                & 14.33                                                & 27.00                                               & 58.67                                                  & 14.00                                                        & 17.00                                                           \\ 
\hline
\textbf{10}                                                & 8.67                                                 & 15.00                                               & 75.33                                                  & 26.33                                                        & 29.00                                                           \\ 
\hline
\textbf{8}                                                 & 4.33                                                 & 11.00                                               & 84.67                                                  & 31.00                                                        & 31.00                                                           \\ 
\hline
\textbf{6}                                                 & 0.00                                                 & 4.67                                                & 95.33                                                  & 43.67                                                        & 43.67                                                           \\
\hlinew{1.2pt}
\end{tabular}\label{tab:snrhuman}
\end{table}

\vfill\eject

\section{Supplementary Evaluation of NI-Occam}\label{app:attempts}

\begin{figure}[t]
  \centering
  \includegraphics[width=0.98\linewidth]{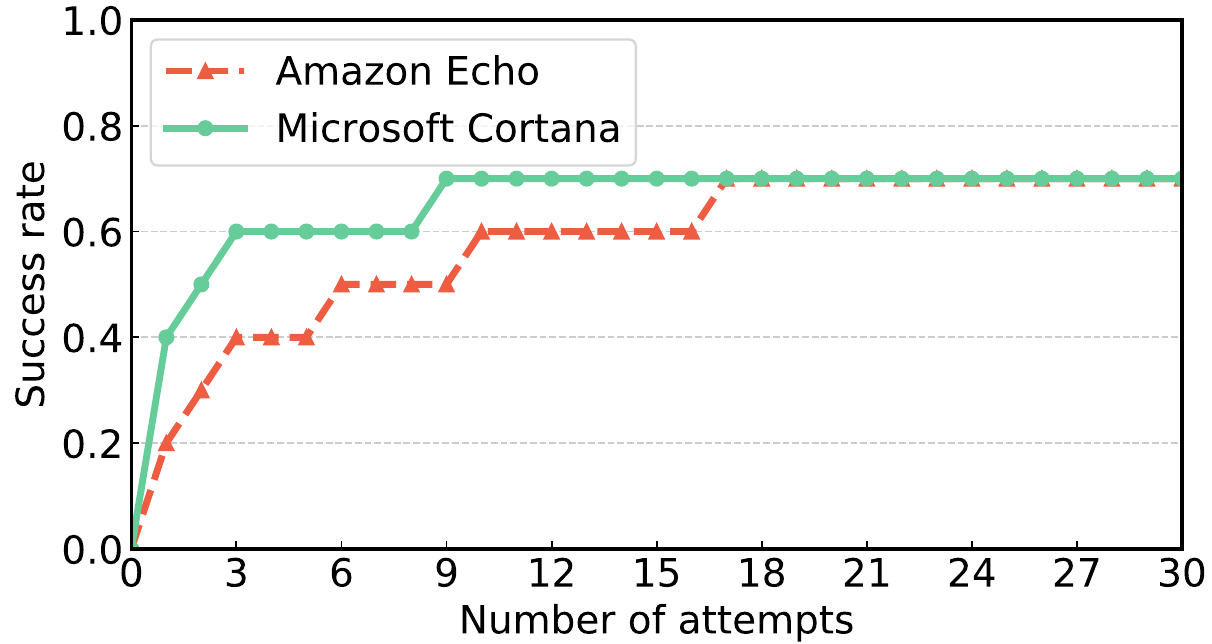} \\
  \caption{\bluej{SRoAs after different numbers of query attempts.}}
  \label{fig:attempt}
\end{figure}

\bluej{As a supplement to Section~\ref{sec:physical}, here we provide the detailed results of our NI-Occam on 10 target phrases in Table~\ref{tab:physicalattack}. Besides, we also conduct an experiment on the impact of the number of attempts.} \bluej{
In previous experiments on NI-Occam, the default setting for the number of query attempts is 3 times, and Devil's Whisper~\cite{Whisper} adopted a number of 30 attempts.}
\bluej{Theoretically, more attempts will lead to higher SRoAs. To illustrate this statement, we conduct an experiment with a range of [1, 30] query attempts using NI-Occam, and the results are given in Figure~\ref{fig:attempt}. 
As shown, at the very start, the increasing number of query attempts can help increase the SRoA. However, when the number of attempts is larger than 17, the SRoA will remain at 70\% and no longer increases, indicating a performance limit.
In practice, however, more attempts also require longer attacking periods, making the attack less stealthy. Hence, we suggest the attacker conduct as few attempts as possible. }

\begin{table*}[p]\small
\centering
\caption{\bluea{Detailed results of NI-Occam against voice control devices.}}\label{tab:physicalattack}

\begin{tabular}{c|c|c|c|c|c|c|c|c|c|c} 
\hlinew{1.2pt}
\multirow{2}{*}{\textbf{Command}} & \multicolumn{2}{c|}{\textbf{Apple Siri}} & \multicolumn{2}{c|}{\textbf{iFlytek}} & \multicolumn{2}{c|}{\textbf{Microsoft Cortana}} & \multicolumn{2}{c|}{\textbf{Google Assistant}} & \multicolumn{2}{c}{\textbf{Amazon Echo}}  \\ 
\cline{2-11}
                                  & \textbf{Attack} & \textbf{SNR (dB)}      & \textbf{Attack} & \textbf{SNR (dB)}   & \textbf{Attack} & \textbf{SNR (dB)}             & \textbf{Attack} & \textbf{SNR (dB)}            & \textbf{Attack} & \textbf{SNR (dB)}        \\ 
\hlinew{1.2pt}
Call my wife             & \checkmark               & 11.57                  & \checkmark               & 7.61                & \checkmark               & 7.61                          & \checkmark               & 11.57                        & \checkmark               & 8.76                     \\ 
\hline
Open the door           &\ding{55}              & -                      & \checkmark               & 8.57                &\ding{55}              & -                             &\ding{55}              & -                            &\ding{55}              & -                        \\ 
\hline
Play music              & \checkmark               & 9.70                   & \checkmark               & 9.70                & \checkmark               & 9.70                          & \checkmark               & 9.70                         & \checkmark               & 11.51                    \\ 
\hline
Send the text            &\ding{55}              & -                      &\ding{55}              & -                   & \checkmark               & 11.51                         &\ding{55}              & -                            &\ding{55}              & -                        \\ 
\hline
{Take a picture}           &\ding{55}              &      -                  &\ding{55}              &      -               &\ding{55}              &       -                        &\ding{55}              &              -                &\ding{55}              & -                        \\ 
\hline
{Turn off the light}       & \checkmark               & 8.61                   & \checkmark               & 8.61                & \checkmark               & 8.61                          & \checkmark               & 8.61                         & \checkmark               & 8.61                     \\ 
\hline
{Open the website}         & \checkmark               & 8.66                   & \checkmark               & 8.50                & \checkmark               & 8.50                          &\ding{55}              & -                            &\ding{55}              & -                        \\ 
\hline
{Make it warmer}           &\ding{55}              &      -                  &\ding{55}              &      -               &\ding{55}              &       -                        &\ding{55}              &              -                &\ding{55}              & -                        \\ 
\hline
{Navigate to my home}      & \checkmark               & 11.54                  & \checkmark               & 11.54               & \checkmark               & 11.54                         & \checkmark               & 11.54                        &\ding{55}              & -                        \\ 
\hline
{Turn on airplane mode}    & \checkmark               & 8.80                   &\ding{55}              & -                   &\ding{55}              & -                             &\ding{55}              & -                            & \checkmark               & 8.80                     \\ 
\hlinew{1.2pt}
\textbf{Average}                  & 6/10            & 9.81                   & 6/10            & 9.09                & 6/10            & 9.58                         & 4/10            & 10.36                        & 4/10            & 9.42                     \\
\hline\hlinew{1.2pt}
\end{tabular}
\vspace{3mm}
\end{table*}

\begin{table*}[!p]
% \centering
\caption{\bluej{Performance of Occam and NI-Occam on large target command sets.}}
\begin{tabular}{c|c|c|c||c|c|c|c} 
\hlinew{1.2pt}
\multirow{2}{*}{\textbf{Commands}} & \textbf{Occam}   & \multicolumn{2}{c||}{\textbf{NI-Occam}} & \multirow{2}{*}{\textbf{Commands}} & \textbf{Occam}   & \multicolumn{2}{c}{\textbf{NI-Occam}}  \\ 
\cline{2-4}\cline{6-8}
                                   & \textbf{Alibaba} & \textbf{Cortana} & \textbf{Siri}       &                                    & \textbf{Alibaba} & \textbf{Cortana} & \textbf{Siri}       \\ 
\hlinew{1.2pt}
watch TV                           & 17.40             & 9.74             & 9.74                & clear notification                 & 9.98             & 8.76             & 8.76                 \\ 
\hline
open the box                       & 11.11            & 8.74             & -                   & when do you get up                 & 10.96            & -                & -                    \\ 
\hline
flip a coin                        & 11.92            & 8.81             & 8.81                & ask me a question                  & 24.97            & 9.18             & -                    \\ 
\hline
set a timer                        & 18.15            & 8.6              & 8.6                 & where is my home                   & 16.21            & 10.23            & 10.23                \\ 
\hline
sing a song                        & 14.26            & 9.29             & 9.29                & where is my hotel                  & 14.42            & 9.73             & -                    \\ 
\hline
crystal ball                       & 9.24             & -                & -                   & mischief managed                   & 20.6             & 8.51             & 8.51                 \\ 
\hline
find a hotel                       & 10.31            & 9.58             & 8.73                & turn off all alarms                & 14.71            & 8.86             & 8.86                 \\ 
\hline
set an alarm                       & 25.01            & 10.09            & 10.09               & turn the volume up                 & 20.37            & 8.28             & 8.28                 \\ 
\hline
surprise me                        & 28.88            & 10.36            & 10.36               & show me the menu                   & 28.69            & 8.61             & 8.61                 \\ 
\hline
take a selfie                      & 17.83            & 8.56             & -                   & where is this place                & 11.52            & 9.27             & -                    \\ 
\hline
reading a book                     & 9.5              & -                & -                   & what is the weather                & 14.09            & -                & 9.03                 \\ 
\hline
close twitter                      & 16.25            & -                & -                   & show me my flights                 & 15               & 9.14             & -                    \\ 
\hline
classical music                    & 14.59            & 8.83             & 9.25                & listen to voice mail               & 17.73            & 7.65             & -                    \\ 
\hline
call my uncle                      & 14.71            & -                & 8.8                 & show me the money                  & 16.83            & 9.11             & 9.11                 \\ 
\hline
tell me a joke                     & 22.6             & 10.27            & 10.27               & show me my alarms                  & 12.78            & 10.12            & 10.12                \\ 
\hline
tell me a story                    & 24.67            & 8.87             & 8.87                & when is my birthday                & 31.49            & 9.28             & 9.28                 \\ 
\hline
turn on the TV                     & 25.07            & 9.15             & 9.15                & turn on the window                 & 10.58            & -                & 8.53                 \\ 
\hline
set a reminder                     & 11.87            & 8.55             & 9.62                & turn off the computer              & 25.2             & -                & -                    \\ 
\hline
clean my room                      & 24.04            & 9.76             & 9.76                & make me a sandwich                 & 19.14            & -                & 7.37                 \\ 
\hline
what time is it                    & 38.07            & 8.84             & 9.41                & turn off the window                & 9.07             & 7.71             & -                    \\ 
\hline
record a video                     & 9.36             & 8.48             & -                   & where is my package                & 20.74            & 8.45             & -                    \\ 
\hline
spin the wheel                     & 18.01            & -                & -                   & turn the volume down               & 21.6             & 9.68             & 9.68                 \\ 
\hline
open instagram                     & 9.71             & 8.59             & -                   & share the new version              & 21.45            & -             & -                    \\ 
\hline
turn on the camera                 & 11.23            & 8.62             & -                   & give me a compliment               & 14.23            & -                & -                    \\ 
\hline
cancel an alarm                    & 13.11            & -                & -                   & show me my messages                & 25.8             & 8.93             & 8.93                 \\ 
\hline
where is my car                    & 22.24            & 8.54             & 9.89                & sing me happy birthday             & 17.09            & 9.34             & 9.34                 \\ 
\hline
how old are you                    & 9.09             & -                & -                   & what is the temperature            & 13.17            & 9.08             & 9.08                 \\ 
\hline
mute the volume                    & 13.99            & -                & 9.32                & turn on the coffee maker           & 20.08            & -                & -                    \\ 
\hline
turn on bluetooth                  & 26.24            & 10.18            & 10.18               & Oscar winner of this year          & 9.85             & -                & -                    \\ 
\hline
turn off bluetooth                 & 15.04            & 9.51             & 9.51                & what would you recommend           & 16.01            & 7.72             & -                    \\
\hlinew{1.2pt}
\end{tabular}\label{tab:large60}

% \vspace{2mm}

\begin{tablenotes}
\item    \footnotesize{
Note that, (i) this table shows the SNRs of the AEs generated from Occam and NI-Occam, and ``-'' denotes this AE fails. 
(ii) In total, the SRoAs of Occam against Alibaba SSR API, NI-Occam against Cortana, and NI-Occam against Apple Siri are 100\%, 71.7\%, and 58.3\%, respectively.
 }
  \end{tablenotes}

% \footnotesize{
% Note that, (1) this table shows the SNRs of the AEs generated from Occam and NI-Occam, and ``-'' denotes this AE fails. 

% (2) In total, the SRoAs of Occam against Alibaba SSR API, NI-Occam against Cortana, and NI-Occam against Apple Siri are 100\%, 71.7\%, and 58.3\%, respectively.
% }
\vspace{5mm}
\end{table*}

\section{Evaluation of Occam and NI-Occam on Large Command Sets}\label{app:large}

As illustrated before, the target phrases in Devil's Whisper~\cite{Whisper} are restricted to several phrases for a substitute model.
Different from Devil's Whisper, Occam is not restricted to the number of targeted phrases, where the adversary can express freely. To illustrate this point, we evaluate Occam~\bluej{and NI-Occam} on 60 targeted phrases. 
In the experiment of Occam, we generate 60 audio AEs with 60 different targeted voice commands by querying Alibaba SSR 30,000 times and feed them into it. The target commands and SNRs of the AEs are presented in Table~\ref{tab:large60}. The 60 AEs have an average SNR of 17.27dB, and all of them can be recognized as the target commands successfully.
\bluej{For NI-Occam, we also generate 60 audio AEs with 60 different voice commands and play the AEs to a Samsung C9 running Cortana and an iPhone 11 running Siri. 
As shown in Table~\ref{tab:large60}, in total, 43 AEs have successfully attacked Cortana,~\ie, the SRoA is 71.7\%. For Siri, there are 35 effective AEs that can be recognized as the target commands,~\ie, the SRoA is 58.3\%.
The results indicate that both Occam and NI-Occam can perform well on large command sets. 
}

\section{Results of Possible Countermeasures}\label{app:Countermeasures}
\bluej{Due to the space limitation, we provide the detailed results of four countermeasures against our attacks in this section. Table~\ref{tab:defense123} shows the  performance of our attacks against different countermeasures, including local smoothing, downsampling and temporal dependency based approach, and Table~\ref{tab:adversarialtraining} presents the performance of our NI-Occam against adversarial training. }

In our experiment of adversarial training, we adversarially train Kaldi (the aforementioned Mini LibriSpeech model) on the Mini LibriSpeech dataset\footnote{https://www.openslr.org/31/}. We set $p$ to $\infty$, iterations to $10$, and the step size $\alpha$ to $\epsilon/5$ for PGD, which are the same with~\cite{DBLP:journals/csl/JatiHPPAN21, joshi2021study}.  Since the prior work~\cite{joshi2021study} demonstrates that adversarial training with $\epsilon > 0.01$ will be not able to converge,  we set $\epsilon$ to $0.002$, $0.004$ and $0.006$. For evaluation, we also generate 10 audio AEs from Mini LibriSpeech model.

\vfill\eject
\begin{table}[t]\small
% \centering
\caption{Performance of our attacks against different countermeasures.}
\begin{tabular}{c|c|c|c} 
\hlinew{1.2pt}
\multirow{2}{*}{\textbf{Countermeasures}}      & \multirow{2}{*}{\textbf{Setting}} & \multicolumn{2}{c}{\textbf{Attack Methods}}  \\ 
\cline{3-4}
                                              &                                   & \textbf{Occam} & \textbf{NI-Occam}            \\ 
\hlinew{1.2pt}
\multirow{2}{*}{\textbf{Local smoothing}}      & $h$ = 1                             & 2/10           & 4/10                         \\ 
\cline{2-4}
                                              & $h$ = 3                             & 0/10           & 4/10                         \\ 
\hline
\multirow{2}{*}{\textbf{Downsampling}}      & DSR = 12kHz             & 0/10           & 3/10                            \\ 
\cline{2-4}

                                              & DSR  = 10kHz                  & 0/10              & 2/10                         \\ 
\hline
\multirow{2}{*}{\textbf{Temporal dependency}} & \multirow{2}{*}{$k$ = Rand(0.2, 0.8)}   & \multicolumn{2}{c}{\textbf{AUC Score}}       \\ 
\cline{3-4}
                                              &                                   & 100\%          & 68\%                         \\
\hlinew{1.2pt}
\end{tabular}\label{tab:defense123}

\begin{tablenotes}
\item    \footnotesize{
Note that in these experiments, we test Occam against Alibaba SSR and NI-Occam against Cortana.  DSR means the downsampling rate. 
 }
  \end{tablenotes}
% \footnotesize{
% Note that, in these experiments, we test Occam against Alibaba SSR and NI-Occam against Cortana.  DSR means the downsampling rate. 
% }
\end{table}

\begin{table}[t]
% \centering
\caption{\bluej{Performance of NI-Occam against adversarial training.}}
\begin{tabular}{c|c|c|c}
\hlinew{1.2pt}
{} &{$\epsilon$} & {WER$^\ddag$} & {SRoA$\dag$}  \\
\hlinew{1.2pt}
Standard training                 & -                                        & 10.69                               & 10/10               \\
\hline
\multirow{3}{*}{Adversarial training$^\$$} & 0.002                                    & 19.82                               & 3/10                \\
\cline{2-4}
                                      & 0.004                                    & 31.54                                 & 1/10                \\
\cline{2-4}
                                      & 0.006                                    & 58.17                                  & 0/10               \\
\hlinew{1.2pt}                                      
\end{tabular}\label{tab:adversarialtraining}

\begin{tablenotes}
 \item   \footnotesize{
Note that, (i) $\ddag$: WER (Word Error Rate) is a common metric to measure the performance of speech recognition systems. WER calculates the ratio of the minimum number of word-level operations, including substitutions, deletions and insertions, required to convert the transcribed text into the target text, to the total number of words. A lower WER indicates a higher accuracy of ASR systems.
(ii) $\$$: Adversarial training is conducted on the Mini LibriSpeech dataset that contains 5 hours of training data and 2 hours of test data.
(iii) $\dag$: Audio AEs are generated using NI-Occam in Kaldi (Mini LibriSpeech model).
 }
  \end{tablenotes}

\end{table}

\end{document}